\documentclass{dukecei}

\usepackage[noend]{algorithm2e} 
\RestyleAlgo{ruled}
\usepackage{threeparttable}     
\usepackage{enumitem}           
\usepackage{comment}            
\usepackage{etoolbox}           
\usepackage[normalem]{ulem}     
\usepackage{pifont}             
\usepackage{cleveref}           

\providecommand{\Description}[1]{}

\makeatletter
\newcommand{\anybreak}[1]{\begingroup\@tfor\@ab@c:=#1\do{\@ab@c\penalty0\relax}\endgroup}
\makeatother

\newbool{inccomment}
\setbool{inccomment}{false}

\newcommand{\jz}[1]{\ifbool{inccomment}{{\color{red}#1}}{}}
\newcommand{\jk}[1]{\ifbool{inccomment}{{\color{teal}#1}}{}}
\newcommand{\qz}[1]{\ifbool{inccomment}{{\color{blue}#1}}{}}

\title{SRE: Accelerating Zero-Knowledge Proofs via \underline{S}calar \underline{R}edundancy \underline{E}limination in Multi-Scalar Multiplication}

\author{
    Jonathan Ku$^{1}$\footnote{Correspondence E-mail: jonathan.ku@duke.edu},
    Guanglei Zhou$^{1}$,
    Jiawen Wu$^{2}$,
    Junyao Zhang$^{1}$,
    Haoxuan Shan$^{1}$,
    Changchun Zhou$^{1}$,\\
    JV Rajendran$^{2}$,
    Yiran Chen$^{1}$,
    Hai ``Helen'' Li$^{1}$\\[1em]
    \normalsize $^{1}$Duke University
    \normalsize $^{2}$Texas A\&M University
}

\begin{document}

\maketitle
\thispagestyle{firstpagestyle} 

\begin{abstract}
Multi-Scalar Multiplication (MSM) is a fundamental cryptographic operation used in Elliptic Curve Cryptography (ECC) across a wide range of applications. Its computation involves intensive elliptic curve point additions (PADD). MSM plays a critical role in cryptographic primitives such as polynomial commitments and zero-knowledge proofs (ZKPs), where it often becomes the runtime bottleneck. Consequently, optimizing MSM is an active area of research focused on reducing computational complexity and accelerating performance. The current state-of-the-art algorithm for large-scale MSM is Pippenger’s bucket method, which consists of three main steps: point merging, bucket aggregation, and window aggregation. Among these, the point merging step is the primary performance bottleneck. Our work targets this major computational bottleneck. To reduce operations for practical applications, we propose a novel Scalar Redundancy Elimination (SRE) algorithm with three approaches: a vanilla array-based approach, a list-based greedy approach, and a list-based exhaustive approach, all incorporating an early stopping technique. Depending on the use case, these strategies can be integrated into our modified MSM algorithm, which is based on Pippenger’s bucket method. As a result, our proposed MSM algorithm, enhanced with the SRE technique, incurs less than 3\% additional search time, while reducing the number of operations by 25\%, leading to a 30\% speedup in runtime compared to the standard Pippenger’s bucket method. This translates to a $10 \sim 20\%$ runtime reduction in various ZKP systems that rely heavily on MSM.
\end{abstract}

\vspace{1em}

\section{Introduction} \label{sec:intro}
Zero-Knowledge Proofs (ZKPs) \cite{zkp} have played a transformative role in modern cryptography, enabling numerous applications since their introduction. Unlike their original formulations, contemporary ZKPs are typically built on Elliptic Curve Cryptography (ECC), with prominent examples including Zero-Knowledge Succinct Non-Interactive Arguments of Knowledge (zk-SNARKs). However, generating such proofs requires significant computation to preserve the property of zero-knowledge. Key cryptographic operations include the Number Theoretic Transform (NTT) and Multi-Scalar Multiplication (MSM) on elliptic curves (ECs), with MSM alone accounting for more than 70\% of the total runtime \cite{cuzk}. Consequently, this work focuses on optimizing MSM to reduce computation time and accelerate zk-SNARKs for practical cryptographic applications.

\begin{figure}[t] 
  \centering
  \includegraphics[width=.7\linewidth]{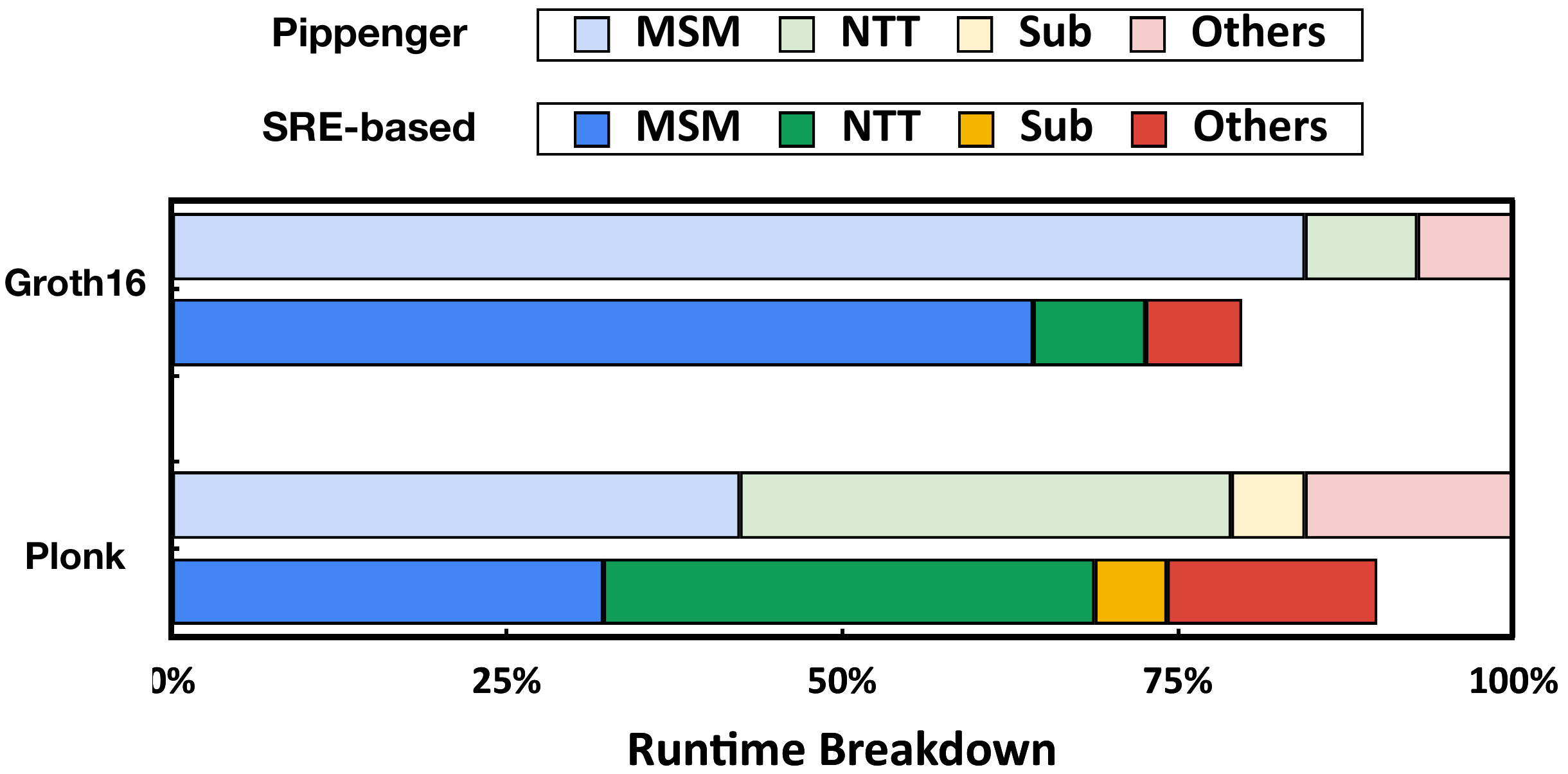}
  \caption{zk-SNARKs runtime breakdown and MSM algorithms speedup comparison.}
  \Description{}
  \label{fig:teaser}
\end{figure}

ECC is one of the most widely used approaches for public-key cryptography. It is based on computations on ECs defined over finite fields. One of its main advantages is the use of smaller key sizes compared to RSA, which has been in use for decades. Due to its numerous benefits, ECC has been widely adopted in various applications, for example, traditional digital signatures, polynomial commitments, and ZKPs mentioned earlier, to name a few. However, computations on ECs still demand significant computing resources due to the large bitwidth operands and the complexity of the underlying cryptographic operations. Among these, MSM on ECs is one of the most computationally intensive tasks. It is a fundamental operation in certain ZKP protocols and accounts for the majority of computation time during the setup phase. As a result, this computational bottleneck hinders the widespread deployment of such applications in practice.  

Joint efforts from both the hardware and software communities have contributed to analyzing \cite{analysis} and optimizing computations, from low-level modular arithmetic \cite{modsram,carrysave,pipemsm}, to EC operations \cite{dbnshardware}, and up to cryptographic accelerators specifically designed for ZKPs \cite{pipemsm,bstmsm,cyclonemsm,hardcamlmsm,msmac,pipezk,zodiacmsm,cuzk,gzkp,elasticmsm}, to make them suitable for real-world applications. However, progress from cryptographic operation perspective has been limited due to the maturity of established approaches. However, operational improvement remains critical, as computational complexity is still a key bottleneck in these applications. In this work, we focus on MSM, which is one of the primary performance bottlenecks in ZKPs. We provide a comprehensive overview of the current methods used in various scenarios along with their respective advantages and limitations.

Pippenger’s bucket method is widely used in state-of-the-art implementations and frameworks for large-scale MSM, such as those employed in ZKP systems \cite{groth16, plonk}. The algorithm merges EC points based on their corresponding scalars. These scalars are split into multiple windows, and matches are identified across these windows to reduce redundant operations that would otherwise occur in a naive computation. We will discuss this algorithm in detail in \Cref{sec:prelim}. Pippenger’s bucket method can be broken down into three main steps: point merging, bucket aggregation, and window aggregation. Among these, point merging accounts for approximately 90\% of the total MSM runtime \cite{gzkp}. To optimize computation time more effectively in ZKP schemes, it is therefore crucial to focus on the point merging step, guided by the well-known Amdahl’s Law \cite{amdahl}. A key insight is that the algorithm can be further improved by identifying and eliminating additional redundancy across multiple windows of scalars, rather than just one. This enhancement allows for greater efficiency beyond what is achieved by the standard Pippenger’s bucket method. The actual runtime breakdown of zk-SNARK applications is shown in \Cref{fig:teaser}, using Groth16 \cite{groth16} and Plonk \cite{plonk} as examples with a problem size of $2^{16}$ under the BN254 EC. The breakdown includes operations such as NTT, MSM, point subtraction (Sub), and others. The results highlight that MSM, and in particular the point merging step, is the primary bottleneck in these cryptographic primitives. 

We propose three approaches for the Scalar Redundancy Elimination (SRE) algorithm: a vanilla array-based method, a list-based greedy method, and a list-based exhaustive method, each combined with an early stopping technique. These approaches offer tradeoffs between search time and the amount of redundancy identified. The early stopping technique aims to determine an optimal threshold, the ratio of matched bits to the total scalar length, beyond which the search yields diminishing returns in redundancy while consuming more time. We integrate SRE into our modified MSM algorithm, based on the Pippenger’s bucket method. Our experimental results, evaluated by various parameters, show that an additional 3\% search cost yields a 25\% reduction in operations, resulting in a general 30\% runtime speedup over the original Pippenger’s method. This demonstrates the effectiveness and efficiency of our approach in practical scenarios. 

As shown in \Cref{fig:teaser}, we compare zk-SNARKs implemented using the standard Pippenger’s bucket method (as the baseline) with those using our proposed algorithm. We perform the analysis using the tool for ZKPs \cite{analysis,analysis_github} for two schemes, Groth16 and Plonk. The tool uses standard Pippenger's bucket method, and we estimate the acceleration through point operation processed. The results indicate potential runtime accelerations of 20.26\% for Groth16 and 9.79\% for Plonk, respectively.

\subsection{Our Contributions}
This work focuses on reducing the computational cost of large-scale MSM using the classical Pippenger’s bucket method, targeting a wide range of applications. We propose a novel method for eliminating redundant operations and demonstrate its potential for integration into both software libraries and hardware accelerators. Overall, our contributions are as follows:

\setlist{nolistsep}
\begin{itemize}[left=0pt]
\item{We provide a comprehensive overview of various cryptographic primitives and applications that utilize MSM on different scales. We analyze state-of-the-art algorithms used in each case, with an in-depth discussion of the specific scenarios we target.}

\item{We propose an innovative approach to tackle large-scale MSM, traditionally performed using Pippenger’s bucket method. We introduce SRE algorithm in three approaches: vanilla array-based, list-based greedy and list-based exhaustive with early stopping technique to search and eliminate redundant operations within the standard method, thereby reducing computations.}

\item{We demonstrated the theoretical redundancy in our scalar analysis and integrated our search method into Pippenger’s bucket method. Our implementation, developed in Python, includes a runtime comparison between different search algorithm approaches. The reduction achieved through the search occurs at the cryptographic operation level, making it compatible with existing acceleration methods used in both software and hardware implementations. Our experimental results show less than a 3\% increase in search time, while achieving a 30\% overall speedup in the MSM computation.}
\end{itemize}

\subsection{Paper Organization}
The remainder of this paper is organized as follows: \Cref{sec:related} reviews related work on accelerating ZKPs from various perspectives. \Cref{sec:prelim} introduces the necessary preliminaries for EC operations and comprehensive overview of MSM. \Cref{sec:sre} describes our proposed SRE algorithm, along with the early stopping technique used to improve performance, illustrated through a simple example. \Cref{sec:proposed} discusses the integration of our proposed MSM algorithm with the SRE algorithm and comparison of memory and storage usage across different methods. \Cref{sec:eval} presents implementation details, evaluation methodology, and our experimental results. \Cref{sec:future} outlines the constraints and future improvement. Finally, \Cref{sec:conclusion} concludes the paper.

\section{Related Work} \label{sec:related}
Algorithms used to compute MSM are chosen based on the scale and the specific application. A more detailed discussion is provided in \Cref{sec:prelim}. In the following, we present related works categorized by method and scale. Since MSM is closely related to scalar multiplication, we also discuss approaches for single-scalar multiplication, as well as MSM techniques ranging from small-scale to large-scale scenarios, highlighting the remaining challenges in each case.

\noindent \textbf{Double/Multi-base number system.}
Double-base number system (DBNS) is a form of representation scheme that uses a base of 2 and 3 to represent a positive integer \cite{dbns}. This can be generalized into 

\begin{equation} \label{eq:dbns}
  k = \sum_{i,j} s_{i,j} 2^i 3^j, \text{with } s_{i,j} \in \{0,1\}
\end{equation}

When the base consists only of 2s, the representation corresponds to a standard binary. DBNS has many applications, largely due to its effect on algorithmic complexity. Some algorithms heavily depend on the number of ones in the representation. As a result, sparse forms with many zeros can be inefficient. In ECs, longer scalar representations lead to more point operations. To mitigate this, EC operations using double base chains, a form of DBNS, have been proposed \cite{dbnsecc,dbsmyu,dbsmmeloni,dbsmsuppa}. Using bases 2 and 3 results in denser scalar forms, improving computational efficiency. This idea also extends to two-scalar multiplication, where shared bases enable further reductions in cost \cite{dbns2sm}.

However, the representation in DBNS is redundant and can yield multiple equivalent forms. While this flexibility is beneficial for two scalars, scaling beyond three becomes challenging. The overhead of searching for optimal double-base representations leads to diminishing returns as the number of scalars grows, posing an open problem for large-scale MSM. Similarly, several works extend DBNS to multi-base number systems (MBNS) to address the difficulty of finding a common radix across multiple numbers \cite{mbsm}. However, MBNS faces the same core limitation: the computational benefit is offset by the increasing cost of finding suitable representations.

\noindent \textbf{Multidimensional scalar multiplication.}
d-MUL \cite{dmul2}, used for multidimensional scalar multiplication, is functionally equivalent to MSM but on a smaller scale. It builds on the addition chain methods described in \cite{dmul1,dmul2,brown}, with various implementation strategies such as differential addition chains and Montgomery ladders. While the approach is promising, it presents several challenges. Scalars are transformed into a state matrix to optimize operations along the addition chain, but continuously updating this matrix is time-consuming and complex. Although optimized output benefits small-scale MSM, the tradeoff becomes clear as the matrix grows at a rate of $N^2$ for N-scalar multiplication, and finding efficient addition sequences becomes harder. As a result, this method is typically practical only when N is between 2 and 8. In zk-SNARKs, where MSM is large-scale, this approach is not feasible. Thus, generalized MSM algorithms are preferred over addition chain-based methods in such settings.

\noindent \textbf{Precomputation.}
To reduce the overall computational cost of MSM, several works \cite{boscostor,fixed,straus} have proposed precomputation-based approaches that aim to minimize the number of operations during execution. These methods involve precomputing a small set of scalar multiples, referred to as base points, which are then used to express every input point as a combination of these precomputed bases. The core idea is derived from the vector addition chain. Computing the shortest addition chain is a well-studied problem, and various techniques have been developed around it. The general strategy involves identifying a minimal addition chain and constructing a corresponding precomputation table to accelerate scalar multiplications. 

Using the Bos-Costor method \cite{boscostor}, MSM can be computed recursively by sorting scalars from largest to smallest as $\mathbf{Q} = (k_0-k_1) \mathbf{P_0} + k_1 (\mathbf{P_0} + \mathbf{P_1}) + ...$. This introduces extra scalar subtractions and point additions, though scalar costs are negligible compared to point operations. The Straus method \cite{straus,fixed}, similar in principle, relies on precomputing $2^{Nc}$ points, where scalars are divided into smaller windows. While breaking MSM into sub-MSMs improves speed, it incurs significant precomputation overhead, impractical at the scales we target (\Cref{sec:prelim}). To balance performance and memory, \cite{variant,fixed} proposes variants based on the Pippenger algorithm, using precomputation tables of varying sizes to reduce repeated aggregation steps. Compared to Straus, these scale better for large N, though all methods still require substantial memory and storage.

\noindent \textbf{Hardware acceleration.}
Efforts to accelerate and optimize ZKP applications have focused on both software and hardware. While MSM is a computational bottleneck in software, it also limits runtime performance in deployed systems. To address this, the community has pursued optimizations across FPGAs \cite{pipemsm,bstmsm,cyclonemsm,hardcamlmsm,msmac}, ASICs \cite{pipezk,zodiacmsm}, and GPUs \cite{cuzk,gzkp,elasticmsm}, all targeting efficient MSM computation under constraints of time, memory, and resources. Some works improve modular multipliers using techniques like the Montgomery transform \cite{montgomery}, Barrett reduction \cite{barrett}, and Karatsuba multiplication \cite{karatsuba}. These low-level optimizations are crucial, as cryptographic primitives ultimately reduce to modular arithmetic operations.

At the EC operation level, performance is improved through dedicated, heavily pipelined hardware that accelerates point additions. These efforts fall under engineering acceleration, relying on the same core operations with minor changes in modular arithmetic or coordinate systems. While affine coordinates \cite{affine} are traditional, projective \cite{projective} and Jacobian coordinates \cite{chudnovsky,jacobian} avoid costly inversions at the expense of additional multiplications. However, these optimizations remain hardware-level and do not introduce algorithmic improvements.

MSM-level optimizations are largely driven by GPU implementations, focusing on parallelization, subtask allocation, and resource sharing to improve throughput. Although effective for large-scale MSM, these approaches use the same underlying algorithm and therefore do not reduce operation counts or algorithmic complexity.

In contrast, this work targets algorithm-level optimization of MSM, specifically Pippenger’s algorithm. We propose the SRE algorithm, which identifies and eliminates redundant point operations during computation. Unlike prior work that focuses on efficient hardware implementation, SRE modifies the algorithm itself. Existing hardware optimization techniques remain applicable, as SRE preserves the same computation dataflow with additional preprocessing for redundancy elimination.
\section{Preliminaries} \label{sec:prelim}
This section provides the background and preliminaries necessary to understand operations on ECs and related cryptographic primitives. A more detailed explanation of the terminology and notations used throughout this paper, particularly in the context of ECC and the computation of MSM, will also be included. We will focus on the broad range of MSM applications and discuss how various methods and algorithms are employed to compute them.

\begin{figure}[b] 
  \centering
  \includegraphics[width=.7\linewidth]{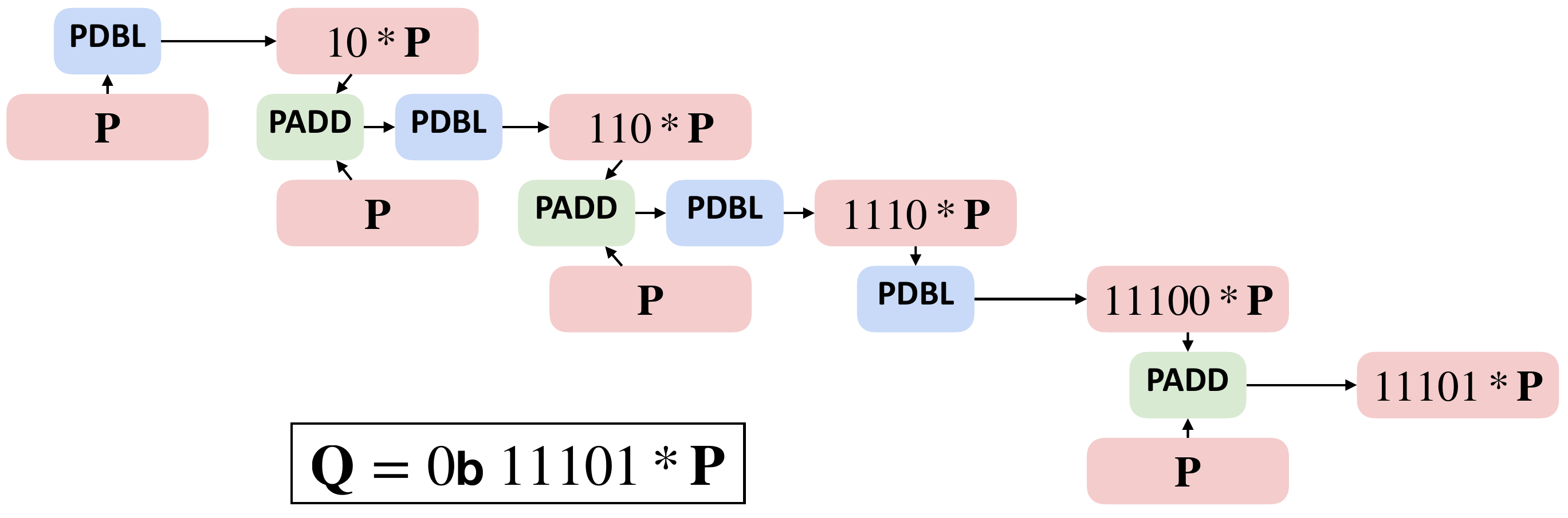}
  \caption{An example of PMUL $\mathbf{Q} = 29*\mathbf{P}$ illustration with double-and-add method.}
  \Description{}
  \label{fig:pmul}
\end{figure}

\begin{table}[t]
  \begin{threeparttable}
  \caption{Summary of elliptic curve parameters. \cite{cyclonemsm,hardcamlmsm,edmsm}} \label{tab:curve}
  \footnotesize
  \setlength{\tabcolsep}{4pt}
  \begin{tabularx}{\linewidth}{c|>{\raggedright\arraybackslash}X}
    \toprule
    BLS12-377 &
    Scalar field $\mathbb{F}_r$ : 253b,
    Base field $\mathbb{F}_q$   : 377b  \\
    \toprule
    r & \ttfamily\anybreak{EqtlXpospVZgtE0eXDewAVmqdv7QAAABChGAAAAAAAE=} \\
    q & \ttfamily\anybreak{Aa46RhfFEOrGOwXAbKFJOxoi2fMA9ROPHvNiL7oJSAAXC11EMAAAAIUIwAAAAAAB} \\
    \midrule
    MNT6-753 &
    Scalar field $\mathbb{F}_r$ : 753b,
    Base field $\mathbb{F}_q$   : 753b \\
    \midrule
    r & \ttfamily\anybreak{AcTGLZLEERAikCLu4s2tt/mXUFuPr+1et+j5bJfYcwf9uSXooO2NmdEk2aFa952xF+d28hgFnbgPDaXLU344aFrM6XZyVKRjiBBxmsQl8OOdVFIs3RGfXpBj3iRegAE=} \\
    q & \ttfamily\anybreak{AcTGLZLEERAikCLu4s2tt/mXUFuPr+1et+j5bJfYcwf9uSXooO2NmdEk2aFa952ybFwoyFmpmz7rypQpISY2ud/5djSZOqTWw4G8PwBXl06gmRcPoTpP2Qd24kAAAAE=} \\
    \bottomrule
  \end{tabularx}
  \begin{tablenotes}
  \item[*] Prime numbers for each field is shown in Base64.
  \end{tablenotes}
  \end{threeparttable}
\end{table}

\subsection{Elliptic Curve Cryptography} \label{subsec:ecc}  
ECC is a form of public key cryptography based on EC over finite fields. The fundamental operations on EC are point addition (PADD) and point doubling (PDBL). PADD adds two distinct points on the curve, while PDBL adds a point to itself, equivalent to multiplying the point by two. These two operations form the basis for all EC computations. Scalar point multiplication (PMUL) multiplies a scalar with a curve point, typically performed using the double-and-add method, as shown in \Cref{fig:pmul}. The average cost of computing one PMUL is $\frac{1}{2}\lambda$ PADDs and $\lambda$ PDBLs, with the worst case requiring $\lambda$ PADDs and $\lambda$ PDBLs, due to a PADD in every iteration, where $\lambda$ is the bitwidth of the scalar. EC operations can be performed in different coordinate systems with varying computational tradeoffs; in this work, we use affine coordinates by default.

ECs are diverse and chosen based on the application. As recommended by NIST, the bit width should be at least 224 bits \cite{nist}. For example, P-256, also known as secp256r1, is widely used in secure communication protocols such as TLS and SSL, as well as in blockchains such as Bitcoin \cite{bitcoin}. Curve25519, also 256 bits, is used in the EC Diffie-Hellman (ECDH) key exchange scheme. For blockchain applications, digital signatures, and ZKPs, pairing-based curves are commonly used. Boneh–Lynn–Shacham (BLS) and Miyaji–Nakabayashi–Takano (MNT) are among the popular choices in zk-SNARKs. In our MSM experiments, we use BLS12-377 and MNT6-753, with scalar field sizes of 253 and 753 bits, respectively. A summary of the EC parameters used in our experiments is shown in \Cref{tab:curve}.

\begin{table}[b]
  \centering
  \caption{Summary of typical N values (number of PMUL).} \label{tab:N}
  \begin{tabular}{c|c}
    \toprule
    Application                           & typical N values\\
    \midrule
    Digital signatures (ECDSA)            & 2\\
    Aggregated signatures (BLS, Schnorr)  & $2^3 \sim 2^{10}$\\
    PCS (KZG, Pedersen)                   & $2^{14} \sim 2^{20}$\\
    zk-SNARKs (Groth16, Bulletproofs ...) & $2^{14} \sim 2^{20}$\\
  \bottomrule
  \end{tabular}
\end{table}

\subsection{Multi-Scalar Multiplication} \label{subsec:msm}
MSM, which is sometimes referred to as multi-dimensional scalar multiplication \cite{mdm}, is an EC operation in which multiple points are multiplied by a scalar through PMUL and summed together through PADDs. The computation is defined as 

\begin{equation} \label{eq:msm}
  \mathbf{Q} = \sum_{i=0}^{N-1} k_i \mathbf{P}_i
\end{equation}

\noindent where k is the scalar defined in the scalar field $\mathbb{F}_r$, $\mathbf{P}$ is a point defined in base field $\mathbb{F}_q$ for a selected EC and N is the total number of point scalar multiplications in MSM. The naive way to compute MSM is by computing PMUL first with double-and-add method and summing them together with PADDs. In this method, it will have the worst case of $N \lambda + \lambda - 1$ PADDs and $N \lambda - N$ PDBLs by the following calculation, where $\lambda$ is the bitwidth of scalars. There are N PMULs and each has at most $\lambda$ PADDs and $ \lambda - 1$ PDBLs as shown in \Cref{fig:pmul} by double-and-add method, the number of operations is N times the above result plus additional $N-1$ PADDs. We will introduce other algorithms to compute MSM in the later section.

MSM is widely used in ECC, appearing in digital signatures and polynomial commitment schemes. The number of scalar multiplications varies by application. For basic digital signatures such as the EC Digital Signature Algorithm (ECDSA) \cite{ecdsa}, the verification step involves two-scalar multiplication. Signature aggregation schemes such as Schnorr and BLS signatures require slightly larger MSMs due to multiple validators or signers, typically ranging from 10 to 1,000. Polynomial commitment schemes (PCS), including KZG and Pedersen \cite{pedersen}, rely on EC operations, making large-scale MSM essential, often involving 1,000 to several million points. In ZKPs such as zk-SNARKs and Bulletproofs \cite{bulletproofs}, most protocols depend heavily on these commitment schemes, where MSM dominates the computation. Examples include Groth16 \cite{groth16} and Plonk (using KZG), as well as Bulletproofs (using Pedersen). Each proof system generally performs 3 to 10 MSMs per instance. A summary of typical N values for various applications is shown in \Cref{tab:N}.

\begin{figure}[b] 
  \centering
  \includegraphics[width=.7\linewidth]{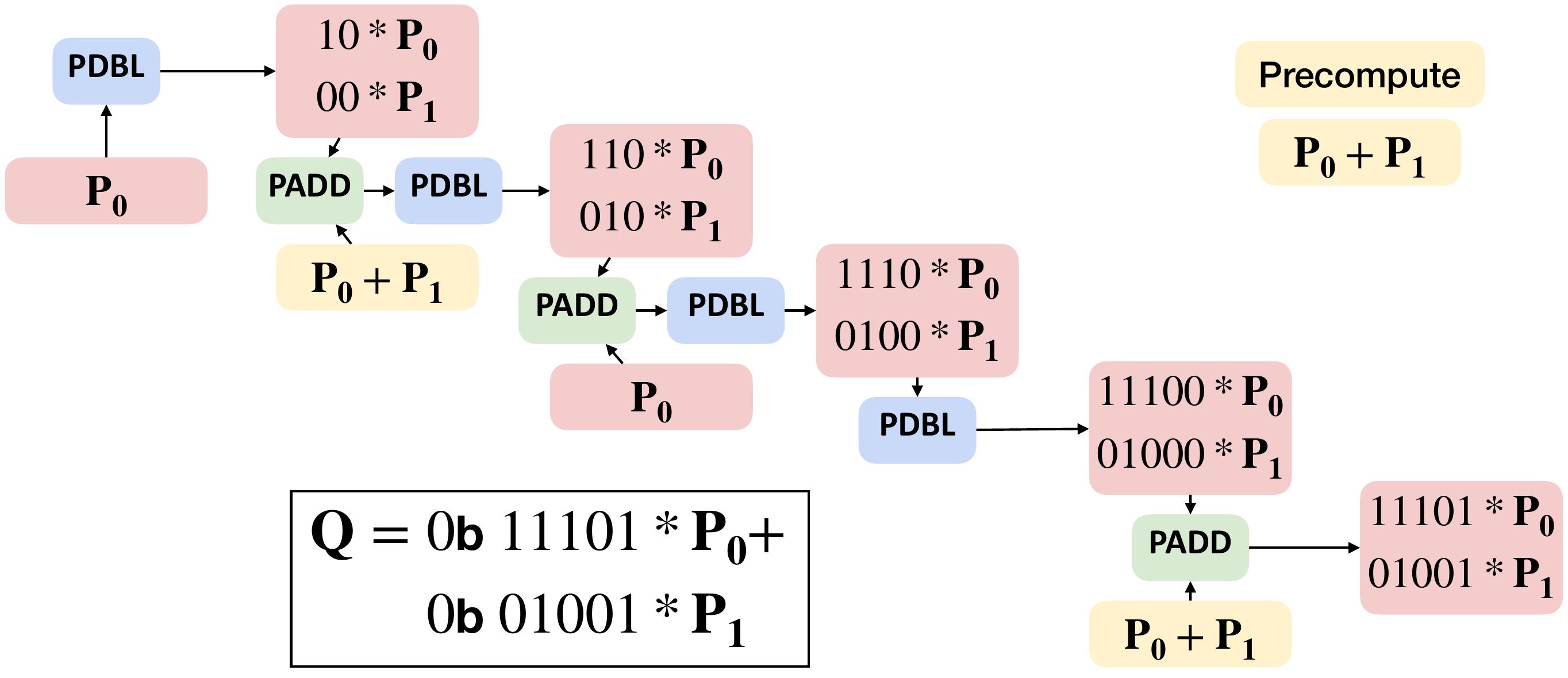}
  \caption{An example of DSM $\mathbf{Q} = 29*\mathbf{P_0} + 9*\mathbf{P_1}$ illustration with Straus-Shamir trick.}
  \Description{}
  \label{fig:shamir}
\end{figure}

\noindent \textbf{Straus-Shamir trick.}
For double-scalar multiplication (DSM), defined as $\mathbf{Q} = k_0 \mathbf{P_0} + k_1 \mathbf{P_1}$, the naive approach computes each scalar multiplication separately using the double-and-add method, then sums the results. This has an average cost of $\lambda$ point additions (PADDs) and $2\lambda$ point doublings (PDBLs). The Straus-Shamir trick \cite{straus,shamir}, also discussed in \cite{trick1,trick2}, offers a simple yet efficient alternative by interleaving the two scalar multiplications. Since individual PMUL results are not required, the trick merges them into a single PMUL using the double-and-add method, aided by precomputing the intermediate point $\mathbf{P} + \mathbf{Q}$. This reduces the cost to roughly one PMUL, with some additional overhead from extra PADDs and precomputation. On average, it requires $\frac{3}{4}\lambda + 1$ PADDs and $\lambda$ PDBLs. An illustration of this trick is shown in \Cref{fig:shamir}.

\noindent \textbf{Pippenger's bucket method.}
Pippenger’s algorithm \cite{pippenger}, revisited in works such as \cite{bernstein,forgery}, was originally developed to accelerate multi-product and multi-exponentiation computations by reducing the total number of multiplications. This concept can be generalized and applied to compute large-scale MSM efficiently using the bucket method, typically for $N = 2^{14} \sim 2^{20}$. The core idea leverages the distributive property of PMUL over PADD on ECs to minimize operations. For example, given two points $\mathbf{P}$ and $\mathbf{Q}$, computing $\mathbf{R} = 2\mathbf{P} + 2\mathbf{Q}$ via the traditional method would involve computing $2\mathbf{P}$ and $2\mathbf{Q}$ separately, followed by summing them, resulting in two PDBLs and one PADD. However, using the distributive law, we can rewrite this as $\mathbf{R} = 2(\mathbf{P} + \mathbf{Q})$, reducing the cost to one PDBL and one PADD. This algorithm, commonly referred to as Pippenger’s algorithm, is widely implemented on hardware platforms for ZKP/MSM acceleration \cite{pipezk,pipemsm,cuzk,gzkp} due to its parallelism and computational efficiency. In the context of MSM, we will refer to it specifically as Pippenger’s bucket method, which can be extended from \Cref{eq:msm} as follows:

\begin{align} \label{eq:pippenger}
  \mathbf{Q} = \sum_{i=0}^{N-1} k_i \mathbf{P}_i 
  &= \sum_{i=0}^{N-1} \left( \sum_{w=0}^{W-1} 2^{wB} k_i[w] \right) \mathbf{P}_i \\ 
  &= \sum_{w=0}^{W-1} 2^{wB} \left( \sum_{i=0}^{N-1} k_i[w] \right) \mathbf{P}_i
\end{align}

\noindent where B is the bucket size in bits, W is the number of windows in a scalar, i.e. $\lceil \frac{\lambda}{B} \rceil$ and $\lambda$ is the bitwidth of the scalar. $k_i[w]$ can be seen as the bucket ranging $0 < k_i[w] < 2^B-1 $ with the total amount of $2^B$ buckets. Each point is placed in one of the buckets and this sub-MSM can be further accelerated due to the sequential increment in the value of the buckets.

\begin{figure}[t] 
  \centering
  \includegraphics[width=\linewidth]{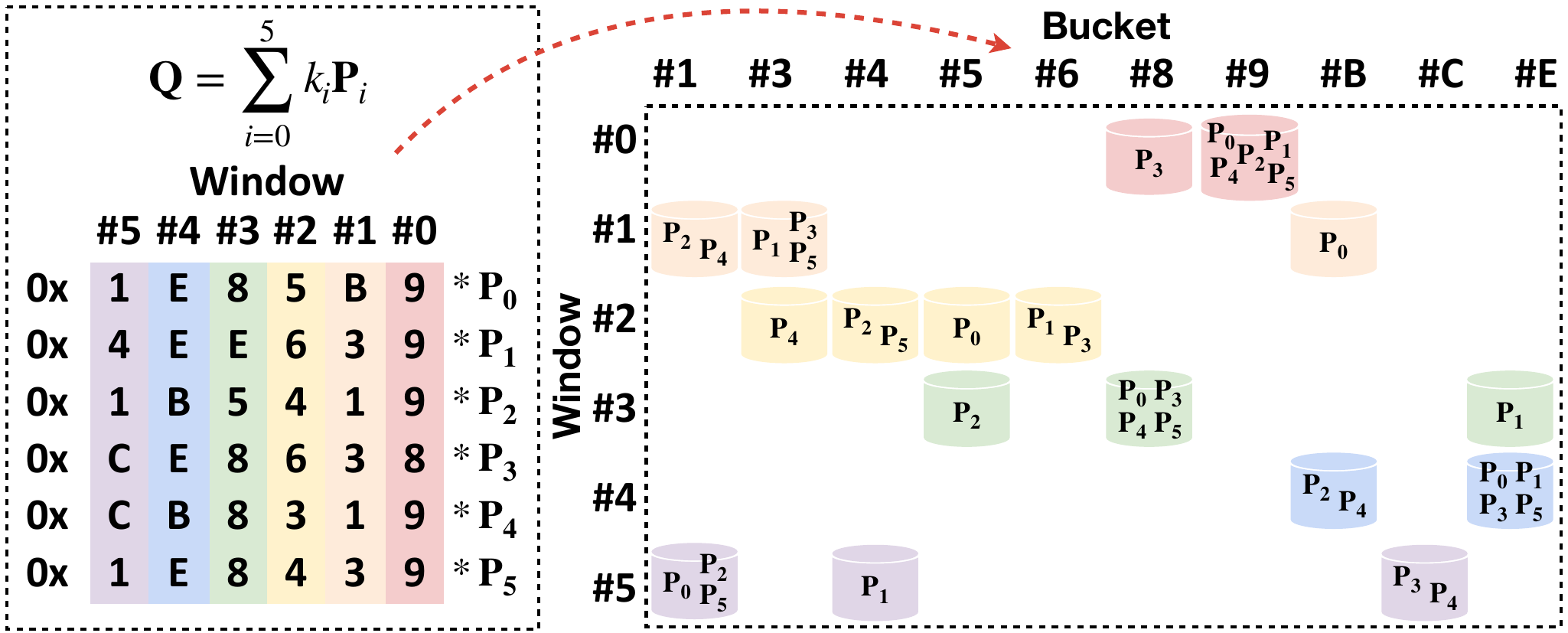}
  \caption{Pippenger's bucket method.}
  \Description{}
  \label{fig:pippenger}
\end{figure}

This is illustrated in \Cref{fig:pippenger}. The toy example demonstrates a 6-scalar multiplication using 24-bit scalars in hexadecimal with a 4-bit window size. Scalars are split into windows shown on the left and in color-coded sequence. Based on the values in the first window, the corresponding points are placed into buckets, as indicated on the right. All points in the same bucket are merged before summing all buckets together. After the first step, a sliding window approach is applied, moving to the next window. By traversing all windows, the scalar multiplication is decomposed into smaller chunks, each associated with a base point, enabling efficient computation using standard PMUL techniques.    

Pippenger's bucket method can be split into three steps: the point merging, bucket aggregation, and window aggregation. Point merging step is illustrated in \Cref{fig:pippenger}. This requires at most $N$ PADDs to place all points in buckets. For bucket aggregation step, it is equivalent to computing

\begin{equation} \label{eq:bucket}
  \mathbf{Q} = \sum_{i=1}^{2^B-1} i \mathbf{B}_i
\end{equation}

\noindent which is a sub-MSM where the scalars are now the bucket values ranging from 1 to $2^B - 1$. It requires $2^{B+1} - 2$ operations using the established algorithm in \cite{cuzk}, which sums points from one new bucket at a time. The window aggregation step is also a sub-MSM, with the number of scalar multiplications equal to the number of windows. As a result, the total number of operations, considering both PADDs and PDBLs, is approximately $W(N + 2^{B+1})$. For large-scale MSM, as shown in \Cref{tab:N}, point merging dominates the computation, accounting for about 90\% of the total. This also shows that the number of PADDs depends on the number of windows, which is proportional to the bitwidth. The equation provides an optimal ratio between $N$ and $B$. However, in practice, the value depends on the user's choices regarding latency and available resources.    

\section{Scalar Redundancy Elimination Algorithm} \label{sec:sre}
\begin{figure*}[t] 
  \begin{subfigure}[t]{.18\linewidth}
    \centering
    \includegraphics[width=\linewidth]{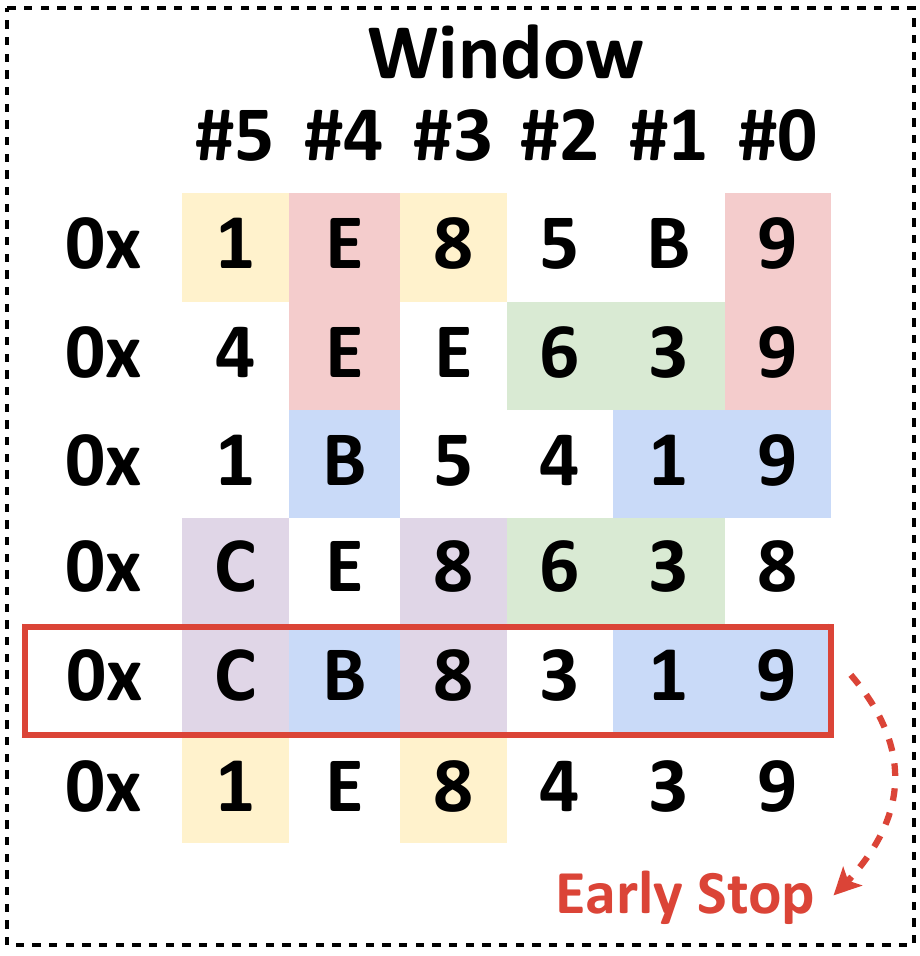}
    \caption{Vanilla array-based.} \label{fig:array}
  \end{subfigure}%
  \begin{subfigure}[t]{.363\linewidth}
    \centering
    \includegraphics[width=\linewidth]{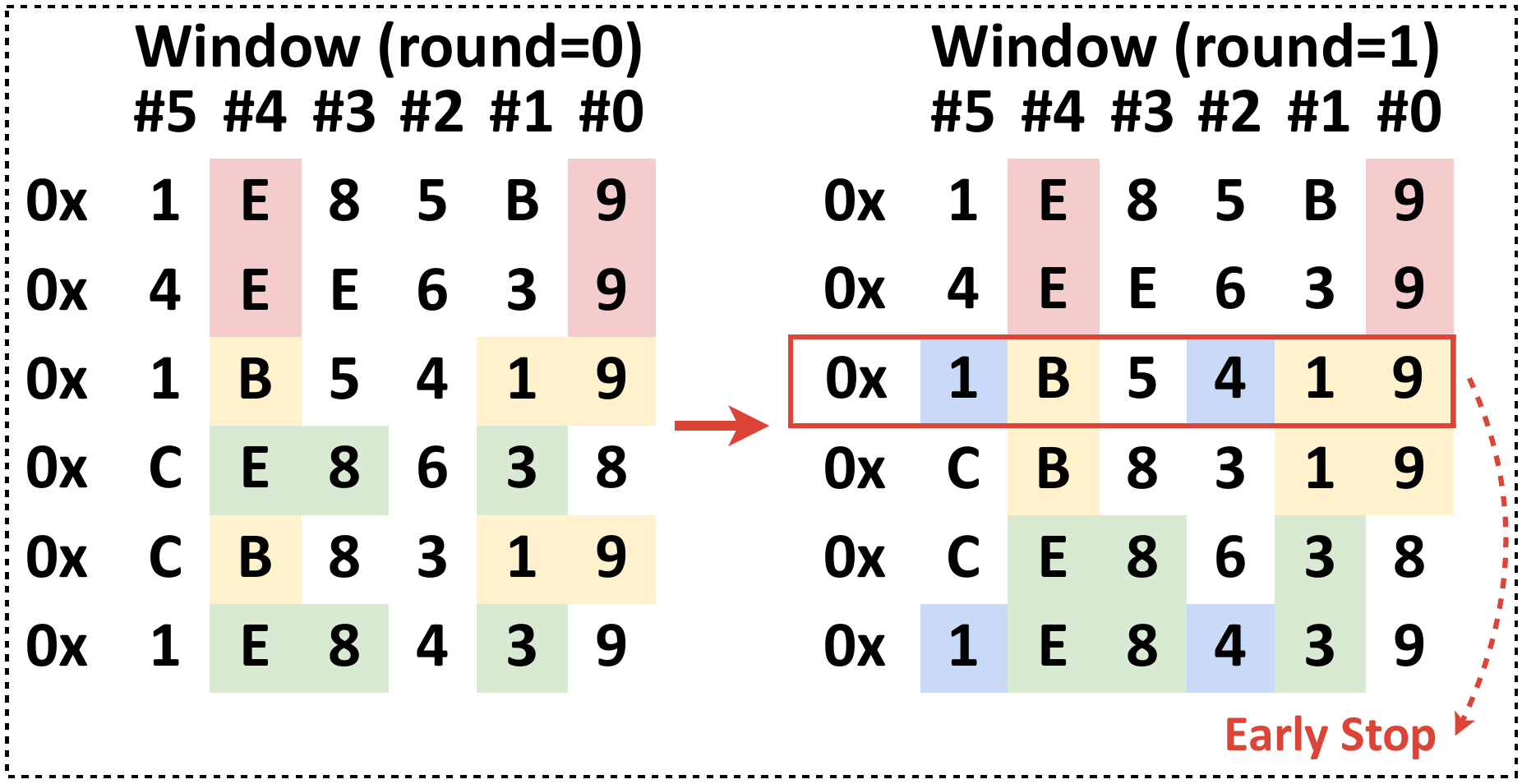}
    \caption{List-based greedy.} \label{fig:greedy}
  \end{subfigure}%
  \begin{subfigure}[t]{.421\linewidth}
    \centering
    \includegraphics[width=\linewidth]{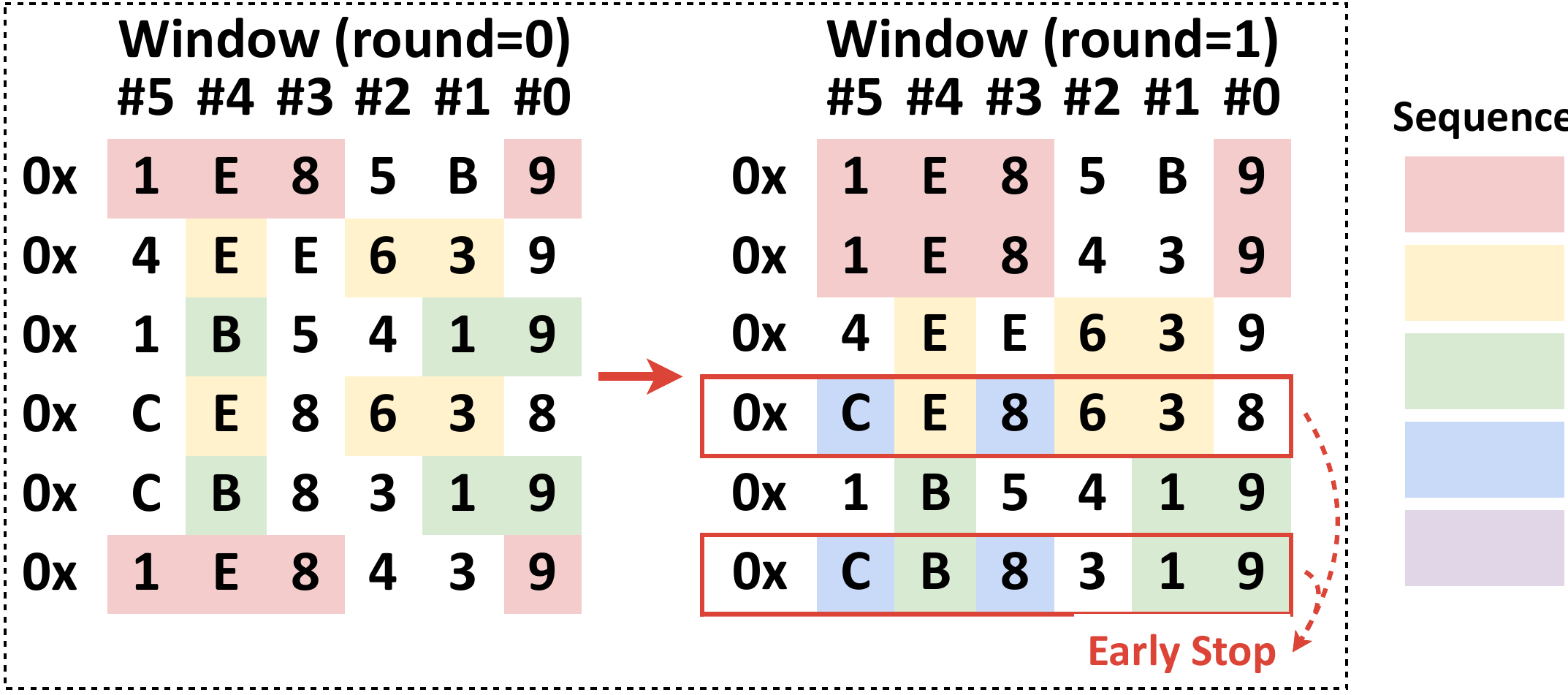}
    \caption{List-based exhaustive.} \label{fig:exhaustive}
  \end{subfigure}
  \caption{Scalar Redundancy Elimination Algorithm in three different approaches with early stopping technique.}
  \Description{}
  \label{fig:sre}
\end{figure*}

As mentioned in \Cref{subsec:msm}, for large-scale MSM, Pippenger’s bucket method reduces the number of operations compared to the naive approach by splitting the scalars into multiple windows and merging multiple points into the same bucket. This optimization reduces the overall computational cost. However, our key observation, despite this algorithm being used for decades, is that the merging step can be extended across multiple windows, rather than handling one window at a time. This idea is inspired by the Straus-Shamir trick in DSM. In that context, instead of precomputing only when both bits are 1, we propose something similar but the other way around, we merge points when both windows are identical. Drawing an analogy from the Straus-Shamir trick: the precomputation phase corresponds to our search algorithm, which acts as a preparatory step for Pippenger’s bucket method. PADDs saved during precomputation represent identical windows that will later be placed in the same bucket at different times. In the Straus-Shamir trick, precomputation can be reused as long as both scalars have 1s in the corresponding positions. Similarly, in our case, redundant operations can be eliminated. An illustration of our observation is provided in \Cref{fig:sre}.

In the classical Pippenger’s bucket method, each window is processed independently without any data dependencies; consequently, all N points are placed into buckets for every window. This approach can be viewed as a vertical split of the MSM, which solves a smaller MSM for each window. In our observation, rather than performing only a vertical split, we first scan the scalars horizontally and then apply the vertical split. In our example, since two points may be placed in the same bucket at different times, the PADD performed for merging is actually redundant. To address this, we propose a Scalar Redundancy Elimination (SRE) prior to executing Pippenger’s bucket method. This approach not only eliminates many redundant PADD operations, but also preserves the structural integrity of the classical Pippenger algorithm. Unlike the method in \cite{fixed}, which relies on precomputing fixed points and requires a large precomputation table, our approach applies to the general case. The only additional overhead introduced by SRE is the preprocessing cost of the search itself. However, as shown in the experimental results later in \Cref{sec:eval}, this search cost is relatively small compared to the total runtime of the MSM computation.

The proposed SRE algorithm has two primary goals: (1) to identify as many merge opportunities as possible within a minimal amount of time, and (2) to find the maximum number of merges without regard to time or resource constraints. To balance between these two extremes, we introduce three intermediate approaches: the vanilla array-based approach, the list-based greedy approach, and the list-based exhaustive approach. In all cases, we focus solely on merging pairs of scalars, although merging multiple scalars simultaneously is theoretically possible.

The following defines the notation and functions used in the algorithm. The input is a group of scalars, which may be structured as either an array or a list, depending on the selected approach. The total number of scalars is denoted by $N$, and the early stopping threshold is denoted by $th$. A detailed explanation of the threshold mechanism is provided in \Cref{subsec:early}. For the list-based approaches, since each scalar is matched only once per round, the total number of merges depends on the number of rounds executed. We determine this based on the convergence behavior of the results. The output is the percentage of redundant operations identified relative to the total number of operations performed during the point merging step. Below is a list of the functions used in \Cref{alg:array,alg:list}:

\setlist{nolistsep}
\begin{itemize}[left=0pt]
\item{EarlyStop(): Determine with a given scalar and threshold, if the amount of merged windows over the whole bitwidth exceeds the threshold or not.}
\item{Cmp(): Compare the two scalars given and return the amount of identical windows to variable $cnt$.}
\item{Replace(): Replace the merged windows with 'X' so that the scalars are not checked repeatedly.}
\item{Len(): Amount of scalars in the group.}
\item{Append(): Append the two scalars into the list.}
\item{Pop(): Pop the two scalars out of the list.}
\end{itemize}

\subsection{Vanilla Array-Based Approach} \label{subsec:array}
To search for scalar merges efficiently, we compare each target scalar with the next scalar in sequence until we find the first one that satisfies the merge condition, namely, that the two scalars share more than one common window. This greedy approach is efficient because it stops at the first valid match.

For example, if two scalars share $c$ identical windows, we can reduce $c$ point additions (PADDs) in the traditional Pippenger’s bucket method to just one PADD. This is because, in the standard method, those $c$ PADDs would occur in different time frames during the point merging step. Once a match is found, the common windows are extracted from both scalars, and the remaining windows are retained for further merging attempts. We scan the scalar list sequentially, attempting to merge each scalar until all of its windows have been processed or the end of the array is reached. This constitutes the first round of the search. In the second round, we begin with the second scalar, and the process continues until all scalars have been traversed. The average-case complexity of this approach starts at $\frac{N}{2}$ comparisons, with a worst-case of $N$, and gradually decreases as scalars become partially merged in subsequent rounds. The algorithm is detailed in \Cref{alg:array}.

\begin{algorithm}[t]
\caption{Vanilla array-based approach} \label{alg:array}
\SetKwData{Scalars}{scalars}
\SetKwData{Threshold}{th}
\SetKwData{Count}{cnt}
\SetKwData{Redundancy}{redundancy}
\SetKwFunction{EarlyStop}{EarlyStop}
\SetKwFunction{Compare}{Cmp}
\SetKwFunction{Replace}{Replace}
\SetKwInOut{Input}{input} 
\SetKwInOut{Output}{output}
\Input{scalars[], N, th}
\Output{redundancy}
\BlankLine
\Redundancy $\leftarrow 0$\;
\For{$i\leftarrow 0$ \KwTo $N -2$}{
  $j\leftarrow i+1$\; 
  \While{$j<N$ \textbf{and} $\sim$\EarlyStop{\Scalars$[i]$, \Threshold}}{
    \Count $\leftarrow$ \Compare{\Scalars$[i,j]$}\;
    \If{\Count $>=2$}{
      \Redundancy $\leftarrow$ \Redundancy $+$ \Count $-1$\;
      \Replace{\Scalars$[i,j]$}\;
    }
    $j\leftarrow j+1$\;
  }
}
\end{algorithm}

\subsection{List-Based Greedy Approach} \label{subsec:greedy}
Similar to the vanilla array-based approach, the list-based greedy approach also uses a greedy algorithm to find the next match, but leverages a different data structure to improve fairness in the merging process. In the array-based approach, scalars are processed sequentially: all windows of the first scalar are merged before moving on to the second. While this is efficient in terms of runtime, it introduces a bias: Scalars appearing earlier in the array have a higher chance of finding matches, while those toward the end have fewer opportunities. This results in a high initial merge rate that decreases as the process progresses.

To address this imbalance, we adopt a list-based structure to store the scalars and modify the traversal strategy. Instead of processing scalars strictly in order, we iteratively scan the entire list in each round to search for merge candidates. When a match is found (i.e., two scalars with more than one common window), both the top scalar and its match are popped from the list and added to a separate container for later processing. The scan then continues with the next scalar now at the top of the list. This strategy ensures that each scalar has a more equal opportunity to participate in a merge during each round, reducing the positional bias present in the array-based method. As a result, the list-based greedy approach typically achieves a higher merge reduction rate. However, the tradeoff is a slight increase in runtime due to the frequent insertions and deletions involved in managing the list.

In terms of complexity, the initial number of comparisons is similar to the array-based method, averaging around $\frac{N}{2}$ in the early rounds. However, the cost quickly decreases as merged scalars are removed from the list, leaving fewer scalars to compare in subsequent iterations. The detailed algorithm is provided in \Cref{alg:list}.

\subsection{List-Based Exhaustive Approach} \label{subsec:exhaust}
Building on the same list-based structure, the exhaustive approach pushes the goal of finding the maximum number of merges, without regard for performance tradeoffs. Instead of stopping at the first valid match, as in the greedy strategy, this method performs an exhaustive search over all remaining scalars to identify the pair with the most common windows. Each merge step is therefore optimal in the sense that it always selects the best possible match available at that time. Once the best pair is found, both scalars are popped from the list and inserted into a new container, just as in the greedy approach. The remainder of the algorithm proceeds in the same manner, repeating this exhaustive matching process in each round until all scalars have been processed.

While this approach shares the same structural properties as the list-based greedy variant, it typically achieves a higher reduction rate due to its thorough search. However, this improvement comes at a higher computational cost. The initial round requires $N$ comparisons, and the number of comparisons decreases by two for each subsequent round as merged scalars are removed from the list. Overall, this method prioritizes maximum redundancy reduction between scalar pairs at the expense of increased runtime and resource usage. The detailed algorithm is presented in \Cref{alg:list}. 

\begin{algorithm}[t]
\caption{List-based approach} \label{alg:list}
\SetKwData{Scalars}{scalars}
\SetKwData{Round}{round}
\SetKwData{Threshold}{th}
\SetKwData{Count}{cnt}
\SetKwData{MaxCount}{maxCnt}
\SetKwData{MaxIndex}{maxIdx}
\SetKwData{Temporary}{tmp}
\SetKwData{Redundancy}{redundancy}
\SetKwData{Approach}{approach}
\SetKwFunction{Length}{Len}
\SetKwFunction{EarlyStop}{EarlyStop}
\SetKwFunction{Compare}{Cmp}
\SetKwFunction{Replace}{Replace}
\SetKwFunction{Append}{Append}
\SetKwFunction{Pop}{Pop}
\SetKwInOut{Input}{input} 
\SetKwInOut{Output}{output}
\Input{scalars[], N, th, round}
\Output{redundancy}
\BlankLine
\Redundancy, \MaxCount $\leftarrow 0$\;
\For{$r\leftarrow 0$ \KwTo \Round}{
  \While{\Length{\Scalars} \textbf{isnot} $0$}{
    \Scalars $\leftarrow$ \Temporary\;
    \For{$i\leftarrow 1$ \KwTo \Length{\Scalars}}{
      \Count $\leftarrow$ \Compare{\Scalars$[i,j]$}\;

      \If{greedy \textbf{and} \Count $>=2$}{
          \MaxCount $\leftarrow$ \Count\;
          \MaxIndex $\leftarrow i$\;
          break\;
      }
      \ElseIf{exhaustive \textbf{and} \Count $>$ \MaxCount}{
          \MaxCount $\leftarrow$ \Count\;
          \MaxIndex $\leftarrow i$\;
      }
      \Redundancy $\leftarrow$ \Redundancy $+$ \MaxCount $-1$\;
      \Replace{\Scalars$[0,$\MaxIndex$]$}\;
      \If{$\sim$\EarlyStop{\Scalars$[0,$\MaxIndex$]$, \Threshold}}{
        \Temporary $\leftarrow$ \Append{\Scalars$[0,$\MaxIndex$]$}\;
      }
      \Pop{\Scalars$[0,$\MaxIndex$]$}\;
    }
  }
}
\end{algorithm}

\subsection{Early Stopping} \label{subsec:early}
The efficiency of SRE comes from identifying the most common windows between pairs of scalars, continuing until the target scalar has no more viable windows to merge. However, regardless of the approach used, each scalar eventually faces diminishing returns, as the likelihood of finding additional merges decreases with fewer remaining windows. To accelerate the search process without significantly compromising operation reduction, we introduce an early stopping technique. The idea is straightforward: we define a threshold ratio based on the number of remaining windows relative to the original bitwidth of the scalar. If the ratio exceeds this threshold, we halt further merging attempts for that scalar, even if additional merges are theoretically possible. At that point, any further merges would require disproportionately more comparisons to achieve the same level of reduction, making them inefficient in practice.

\subsection{SRE Illustration} \label{subsec:sreill}
\Cref{fig:sre} illustrates our proposed SRE algorithm using all three approaches with and without the early stopping technique. This example uses six 24-bit scalars represented in hexadecimal form, with a window size of 4 bits. The scalars are identical to those used in the example shown in \Cref{fig:pippenger}, allowing for a direct comparison. In the vanilla array-based approach, the array structure remains unchanged throughout the search process. Therefore, it is represented using a single table. In contrast, the list-based greedy and exhaustive approaches involve dynamic modifications to the list structure. Elements are appended to or popped from a new list in each round. As a result, in our example with two rounds, they are represented by two tables.

The table is interpreted as follows. In \Cref{fig:array}, we begin by comparing the first scalar with the second scalar. A match is found and the matching windows are highlighted with a red box. The search then continues by comparing the first scalar with the third, fourth, and fifth scalar, but no additional matches are found. When the first scalar is compared with the sixth scalar, another match is found, highlighted with a yellow box. The process then moves on to the second scalar and continues in sequence until all scalars have been traversed and no more matches remain. Each matched window is marked with a colored box, and the sequence of comparisons is represented using a color legend shown on the right. For the list-based approaches, each time a match is found, both matching scalars are popped from the original list and appended to a new list, which gradually creates a more structured and organized result. The next scalar to be processed becomes the new head of the list. In the case of early stopping, consider an 80\% threshold, meaning we stop searching once a scalar has merged 20 out of 24 bits, i.e., $\frac{20b}{24b} = 83\% > 80\%$. If a scalar exceeds this threshold, it is considered fully searched and is excluded from further comparison, even if some potential merges remain.

\section{Proposed SRE-based MSM Algorithm} \label{sec:proposed}

Our proposed MSM algorithm builds upon Pippenger’s bucket method by introducing the SRE prior to placing points into buckets. This modification improves the efficiency of the original Pippenger’s approach. The SRE algorithm, which searches scalar windows to reveal computational redundancies, has been extensively discussed in \Cref{sec:sre} through several different strategies and approaches. In this section, we present the complete MSM algorithm that seamlessly integrates SRE into Pippenger’s bucket method. As illustrated in \Cref{fig:pippenger}, the original bucket method employs a sliding window-based technique to determine how points are assigned to individual buckets. This process is visually represented using colored sequences, where the points are traversed in a vertical manner, starting from the topmost scalar and moving downward. In contrast, our proposed SRE-based algorithm alters the traversal order of the scalars. Specifically, we first scan the scalars horizontally, from left to right, before moving vertically through each column. This traversal pattern enables the detection of identical window patterns between pairs of scalars. Once a matching pair is identified, the corresponding points are combined using a PADD, resulting in a single merged point. This newly formed point is then placed into the appropriate bucket based on its window value. Because the merging is performed before the placement of points into buckets, each successful match allows us to save one PADD operation compared to the original Pippenger’s bucket method. Thus, every time two scalars share an identical window and are merged early, the total number of additions required by the MSM process is reduced. This elimination of operations contributes to the improved performance and computational efficiency of our proposed algorithm.

Due to the nature of our proposed algorithm, which aims to identify and eliminate redundant operations, it remains fully compatible with any acceleration techniques used in hardware or software implementations. As noted in \cite{fixed}, a point addition has nearly the same computational cost as a point subtraction, since subtraction can be achieved simply by negating the y-coordinate of the point. This negation step incurs minimal overhead compared to other modular arithmetic operations. Leveraging this property, the traditional Pippenger’s bucket method, as shown in \Cref{eq:pippenger}, can reduce the bucket size by half, allowing the bucket values to range within $0 < k_i[w] < 2^{B-1} $, and the window values within $-2^{B-1} < k_i[w] < 2^{B-1} $. This transformation is accomplished by converting the scalars in MSM into their non-adjacent form (NAF). Our SRE algorithm can still be applied prior to executing this variant of Pippenger’s method, as redundancies remain even when the scalar bits have the value -1. 

With regard to alternative coordinate systems and various modular multipliers, as thoroughly discussed in \Cref{sec:related}, these techniques focus on accelerating computation below the level of EC operations, and thus do not interfere with the integration of our proposed algorithm into existing computing frameworks. However, to clearly demonstrate computational savings at the algorithmic level, we will compare our proposed method directly with the standard Pippenger’s bucket method, excluding the influence of other acceleration techniques.  

\begin{figure}[b] 
  \centering
  \includegraphics[width=\linewidth]{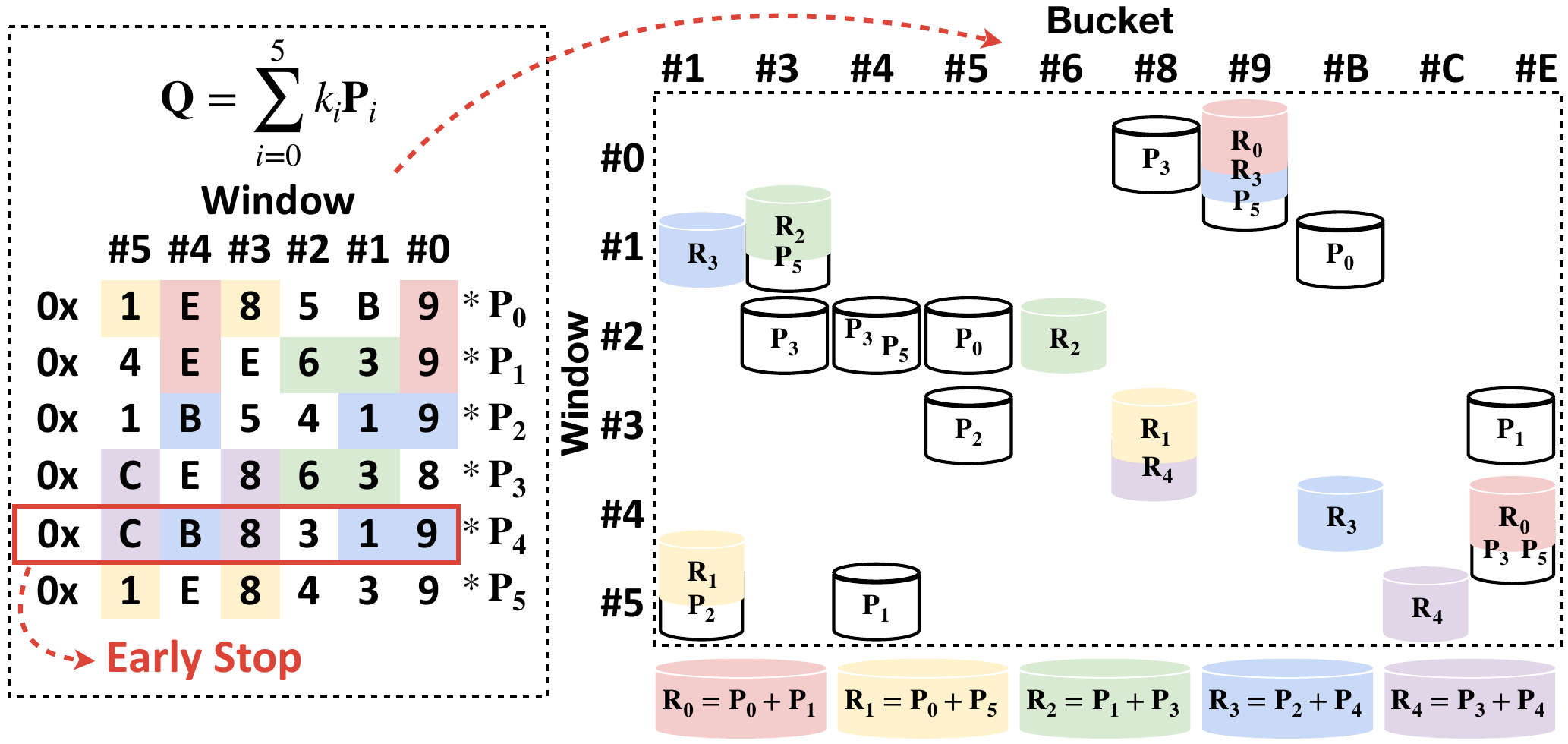}
  \caption{Proposed algorithm with SRE using vanilla array-based approach.}
  \Description{}
  \label{fig:proposed}
\end{figure}

\subsection{Algorithm Illustration} \label{subsec:ill}
\Cref{fig:proposed} illustrates our proposed SRE-based MSM algorithm, which is based on Pippenger’s bucket method. The toy example is identical to the one used in the original Pippenger’s bucket method shown in \Cref{fig:pippenger}, allowing a direct comparison. In our algorithm, rather than immediately splitting windows and placing points into buckets, we first apply the SRE algorithm and perform point additions beforehand. These additions are done prior to point placement because the resulting merged points may belong to multiple buckets. This step marks a key distinction between the traditional Pippenger’s method and our approach: avoidance of redundant computations that would otherwise occur when the same points are used multiple times. The resulting points from this early merging process are shown at the bottom of the figure as  $\mathbf{R_i}$, with the colors indicating the search sequence, as used in \Cref{fig:sre}.

In this example, we utilize our vanilla array-based approach as the SRE algorithm; however, this can be replaced with alternative SRE strategies as needed. The remaining unmatched windows, those for which no redundancy is detected, are processed using the traditional point placement strategy, depicted with black-bordered buckets. In scenarios where early stopping is applied, we can simply proceed by placing the unmatched points into buckets and terminating further searching for that specific point, thus optimizing computation for that case.

\begin{table}[t]
  \centering
  \caption{Memory \& storage requirement for each method.} \label{tab:storage}
  \begin{threeparttable}
  
  \begin{tabular}{c|c}
    \toprule
    Method                              & Memory \& storage\\
    \midrule
    Naive method                        & $N\cdot\mathbf{P}        $\\
    Pippenger's bucket \cite{pippenger} & $(N+2^B)\cdot\mathbf{P}  $\\
    Straus \cite{straus}                & $N2^B\cdot\mathbf{P}     $\\
    TCHES'23 \cite{fixed}               & $3NW\cdot\mathbf{P}      $\\
    SRE-based                           & $(N+2^BW)\cdot\mathbf{P} $\\
  \bottomrule
  \end{tabular}
      \begin{tablenotes}
      \scriptsize
      \item[*] Typical values are $N=2^{6} \sim 2^{16}, B=2^{2} \sim 2^{3}, W=2^{5} \sim 2^{7}$
    \end{tablenotes} 
  \end{threeparttable}
\end{table}

Comparing \Cref{fig:pippenger,fig:proposed}, we observe another major difference in terms of overall resource usage. In Pippenger’s bucket method, the placement of points across windows is uncorrelated. As a result, only $2^B$ buckets are needed at any given time, where $B$ denotes the window size in bits. In contrast, our method performs redundancy checks across all windows, making additional memory allocation for intermediate results and reserving buckets across different windows unavoidable. Consequently, the total number of buckets required by our algorithm is $2^B \times W$, where $W$ represents the number of windows in a scalar. Despite this increased memory usage, our method does not require storage of any precomputed EC points, unlike approaches such as \cite{boscostor,fixed,straus}, which involve both extensive precomputation and significantly larger storage requirements to hold look-up tables (LUTs), in addition to the memory used for computation. A detailed comparison of memory and storage usage is provided in \Cref{tab:storage}. In this theoretical comparison, memory refers to computational resources used for storing intermediates during execution, whereas storage denotes data that remain stationary throughout the computation. This includes all scalars and points used in MSM, as well as precomputed lookup tables, since they remain unchanged during the MSM process.

\section{Implementation \& Evaluation} \label{sec:eval}
Implementation details and evaluation methodology are discussed in depth in this section. We implemented our proposed SRE algorithm using three different approaches and integrated it into Pippenger’s bucket method to compute MSM. First, we analyzed the potential and theoretical redundancy uncovered by our search algorithm. Second, we compared the integrated version of the Pippenger’s bucket method with both the traditional naive method and the original Pippenger’s bucket method in terms of runtime, to assess how the reduction in operations translates into actual runtime improvement, taking into account the additional cost introduced by the search process.   

\noindent \textbf{Software.} 
We implemented our algorithms and conducted experiments using Python. For our SRE algorithm, we generated $N$ random scalars in the scalar field $\mathbb{F}_r$ that correspond to a given EC. For MSM, we generated $N$ random points in the base field $\mathbb{F}_q$ using the curve generator. Our implementation supports various ECs, different numbers of scalar multiplications, and window sizes of 4 and 8 bits. We used the BLS12-377 curve and the MNT6-753 curve to evaluate our algorithm under 256-bit and 756-bit scalar fields, respectively. The parameters for each curve are listed in \Cref{tab:curve}. We utilized ECPy \cite{ecpy}, an open-source EC library that provides point operations, for functional correctness checks.

\noindent \textbf{Hardware.} 
All of our experiments were conducted on a machine equipped with an Intel(R) Xeon(R) Gold 6254 72-core CPU running at 3.10 GHz, along with 1.58 TB of RAM. Our implementation is not parallelized and therefore utilizes only a single CPU core. We report the average runtime over 10 executions.

\begin{figure*}[t] 
  \begin{subfigure}[t]{.422\linewidth}
    \centering
    \includegraphics[width=\linewidth]{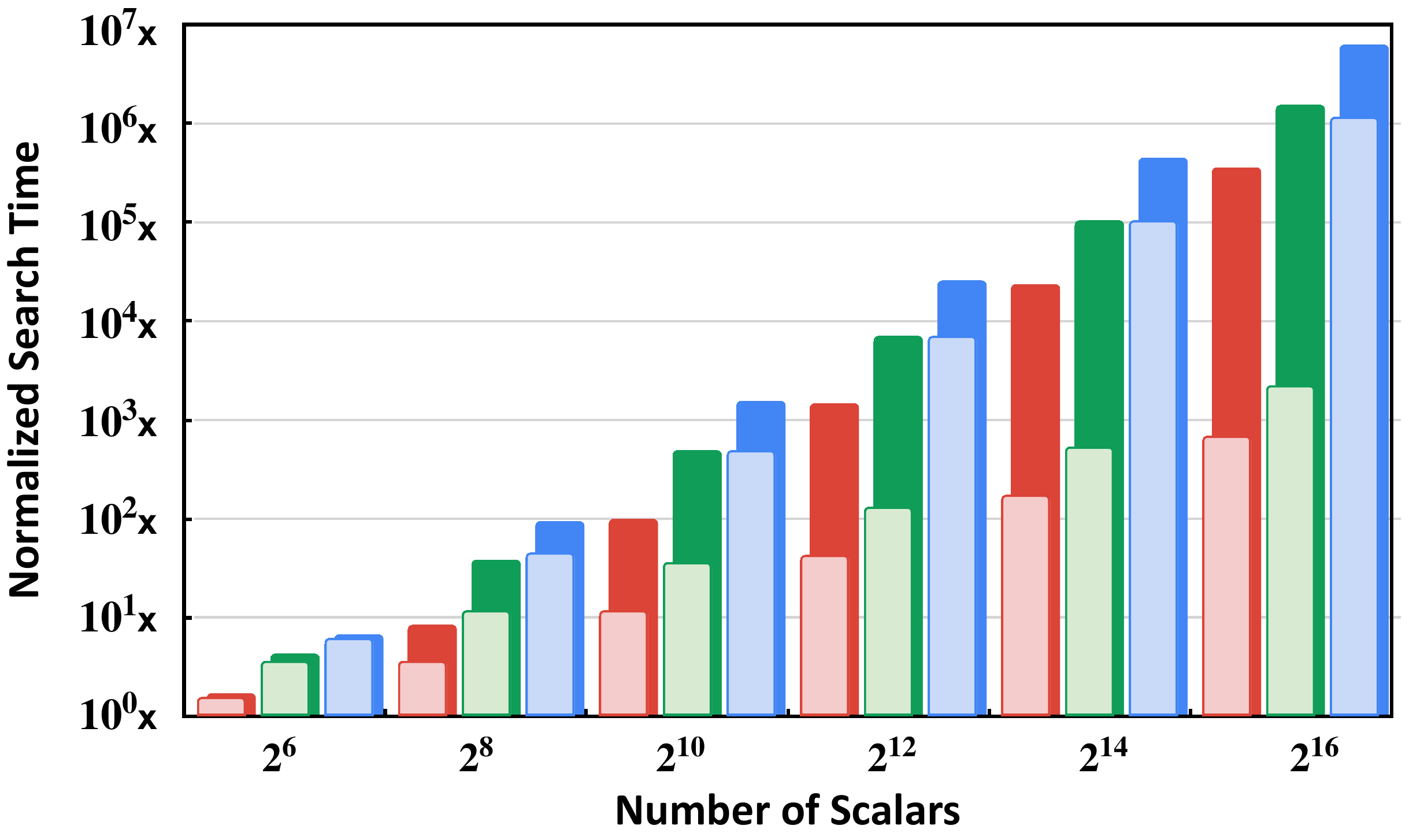}
    \caption{Search Time vs. Number of Scalars.} \label{fig:sre_N_t}
  \end{subfigure}%
  \hfill
  \begin{subfigure}[t]{.411\linewidth}
    \centering
    \includegraphics[width=\linewidth]{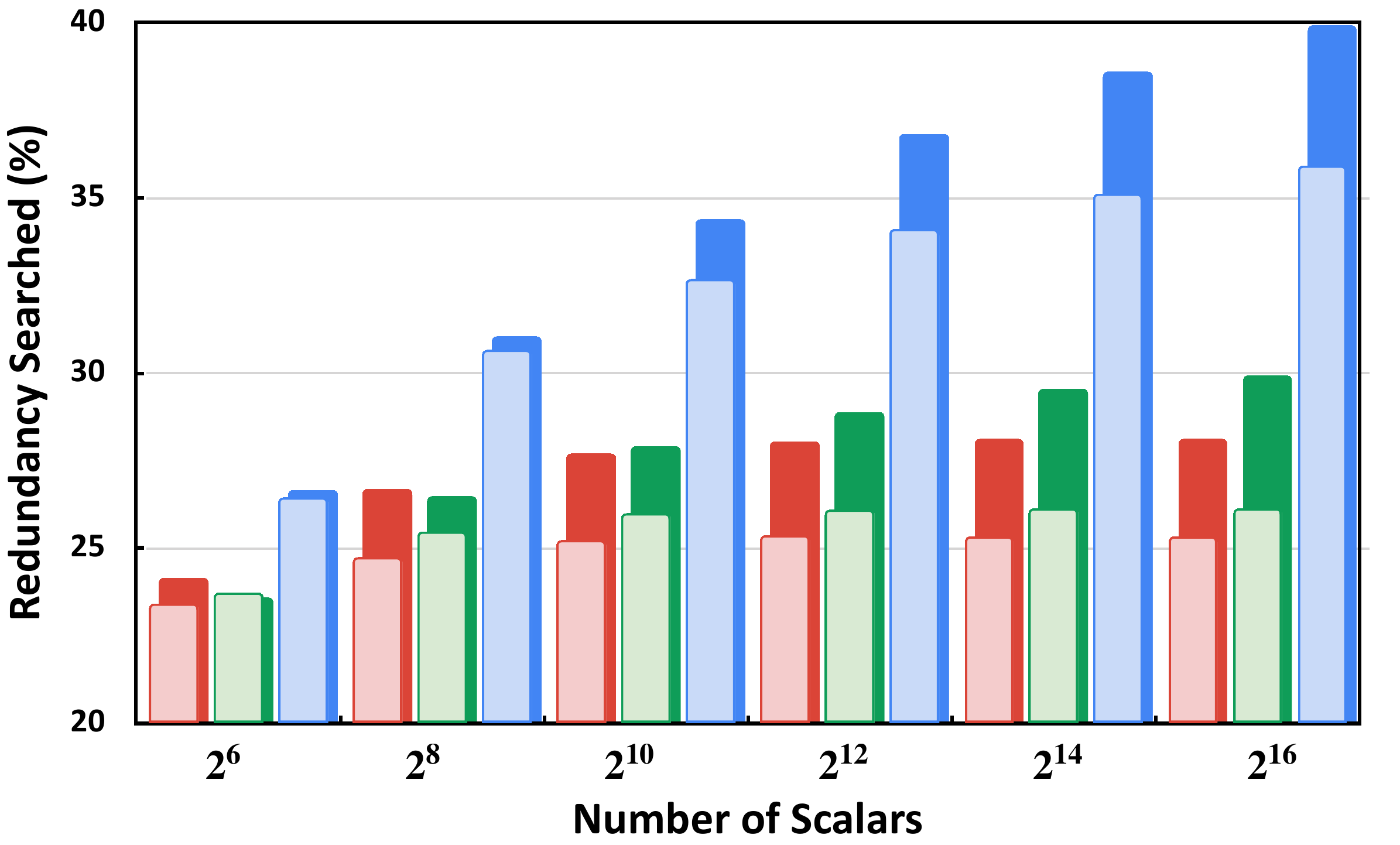}
    \caption{Redundancy Searched vs. Number of Scalars.} \label{fig:sre_N_pct}
  \end{subfigure}%
  \begin{subfigure}[t]{.15\linewidth}
    \centering
    \includegraphics[width=\linewidth]{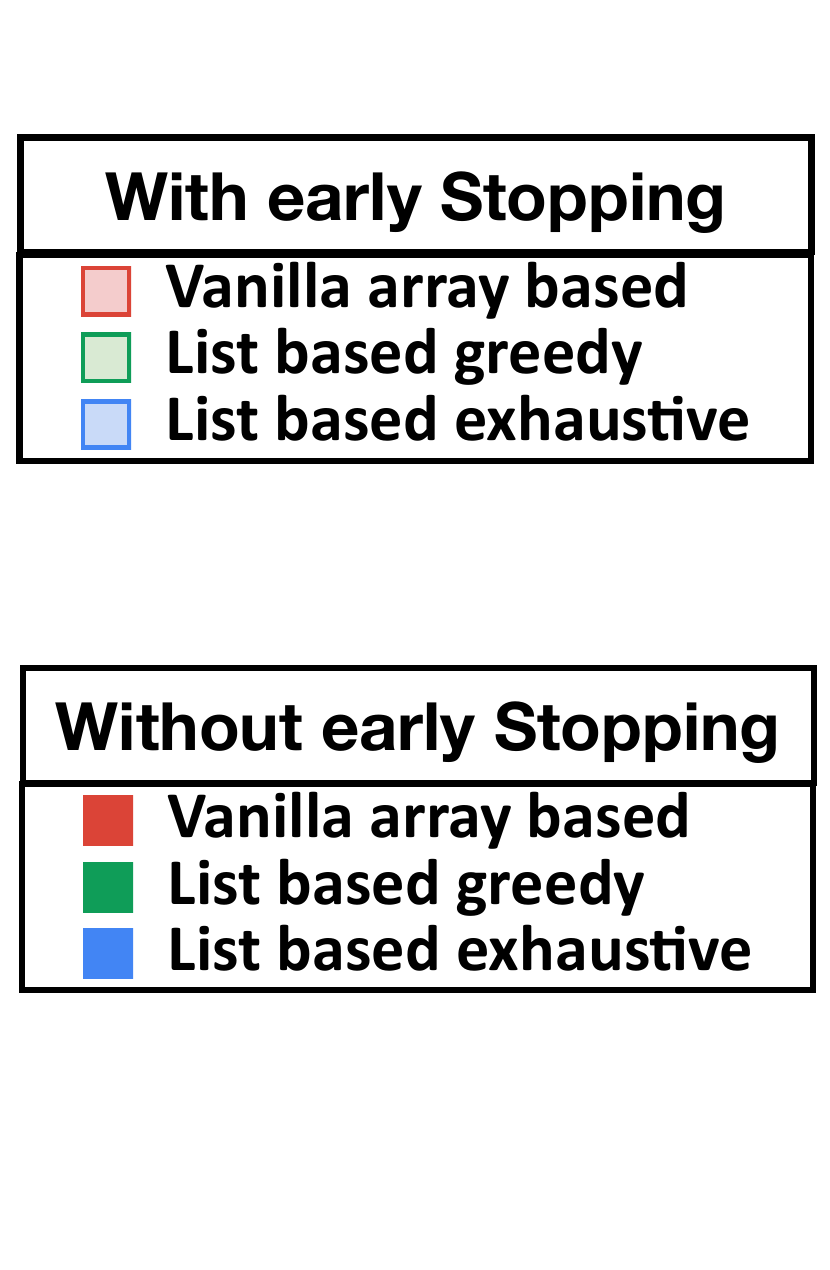}
  \end{subfigure}
  \caption{Scalar Redundancy Elimination Algorithm: Search Time, Redundancy vs. Number of Scalars in three approaches.}
  \Description{}
  \label{fig:eval1}
\end{figure*}

\subsection{Scalar Redundancy Elimination Analysis}
\noindent \textbf{SRE algorithm \& search time.} 
In this section, we evaluate our proposed SRE algorithm using different approaches, both with and without the early stopping technique. In our first experiment, we compare the average runtime of the SRE algorithm for varying numbers of scalars, ranging from $2^6$ to $2^{16}$, using 256-bit scalars and a window size of 4 bits. The threshold used in this experiment is 80\%, meaning that once more than $(256 / 4) \times 0.8 = 51.2$ windows, or 204.8 bits, are matched, the scalar is considered complete and the search is terminated early. We use the search time for the $2^4$ scalars with the vanilla array-based approach as a baseline to normalize the search time in the graph. The normalized search time is presented on a logarithmic scale, where the results with and without the early stopping technique are overlaid to highlight the additional runtime cost incurred in exchange for a more exhaustive search.

As discussed in \Cref{sec:sre}, the three approaches explore redundancy under different resource constraints. In the vanilla array-based approach, scalars are stored in an array and scanned sequentially, with minimal data movement; thus, the search time is relatively low. In contrast, the two list-based approaches involve appending and popping scalars, which introduce additional overhead. The difference in search time between these two approaches stems from how scalars are selected for comparison. In the greedy approach, the algorithm selects the next available scalars that potentially match, resulting in shorter traversal time. Meanwhile, the exhaustive approach evaluates all scalars in the list before selecting the best match for a given target scalar, leading to longer runtimes. Our experimental results in \Cref{fig:eval1} align with these observations.

For different numbers of scalars, the search time increases exponentially regardless of the approach used. However, as shown in \Cref{fig:sre_N_t}, the growth rate is significantly lower when the early stopping technique is applied. This is because early stopping allows the search to terminate once the threshold is reached, thereby avoiding unnecessary computations with diminishing returns. Since both the vanilla array-based and the list-based greedy approaches use greedy algorithms, their search-time growth rates are similar. The vanilla array-based approach is slightly faster due to the more efficient data structure. As the number of scalars increases, the benefits of early stopping become more pronounced. When the number of scalars is less than 1,000, the difference in runtime with and without early stopping is relatively small. However, beyond 1,000 scalars, the runtime nearly doubles, for example, in the case of $2^{16}$ scalars. For the exhaustive approach, the search time grows at a higher rate compared to the other two methods, though the relative improvement from early stopping is smaller. This is because a large portion of the total search time is already consumed in scanning all candidates, making early termination less impactful.

Each experiment was executed ten times, and the results reflect the average runtime. In general, the standard deviation for all runs is below 10\%, with the exception of a few outliers, particularly in the exhaustive search and for $2^{16}$ scalars, due to the significantly longer runtimes. For these large-scale cases, the experiments were performed only once, as they took several orders of magnitude longer to complete, up to a million times longer than the baseline with $2^4$ scalars.

\noindent \textbf{SRE algorithm \& redundancy searched.}
Given the normalized search time results in \Cref{fig:sre_N_t}, we next evaluate whether the additional time cost translates effectively into redundancy gains. \Cref{fig:sre_N_pct} shows the percentage of redundancy found relative to the point merging step in the Pippenger bucket method. As discussed in \Cref{sec:intro}, point merging is the primary bottleneck in the Pippenger method. Therefore, the redundancy identified by our approach can directly translate to a reduction in MSM computation time. In our second experiment, we compare the redundancy discovered by the SRE algorithm for different numbers of scalars, using the same parameters described previously.

In the case of the vanilla array-based approach with the early stopping technique, the redundancy ratio remains nearly constant even as the number of scalars increases. This makes it the most efficient approach, as it achieves the smallest increase in search time. With early stopping, the redundancy reaches a plateau because scalars terminate once the threshold is met, indicating that all scalars are reaching the threshold and no further search is performed. Even without early stopping, the increase in redundancy remains relatively limited. For the list-based greedy approach, the difference in performance compared to the array-based approach is minimal when the number of scalars is small. However, as the number increases, this method yields slightly higher redundancy percentages than the array-based version, though at the cost of significantly greater search time. This difference becomes more pronounced when early stopping is not applied. Lastly, we consider the list-based exhaustive approach. As expected, since it performs an exhaustive search, it achieves the highest redundancy among the three methods. Redundancy continues to increase with the number of scalars, both with and without early stopping. This represents the best-case scenario in terms of redundancy, assuming no resource or time constraints. However, as shown in \Cref{fig:sre_N_t,fig:sre_N_pct}, this comes at the cost of a significantly longer search time, often times by orders of magnitude greater than the other methods.

All three approaches achieve over a 20\% reduction in operations compared to using only the Pippenger’s bucket method in theory. Nevertheless, considering practical runtime performance in software, we use the vanilla array-based approach in later experiments. Although all three approaches demonstrate substantial redundancy improvements, the array-based method offers the best tradeoff between efficiency and performance.

\begin{figure*}[t] 
  \begin{subfigure}[t]{.5\linewidth}
    \centering
    \includegraphics[width=\linewidth]{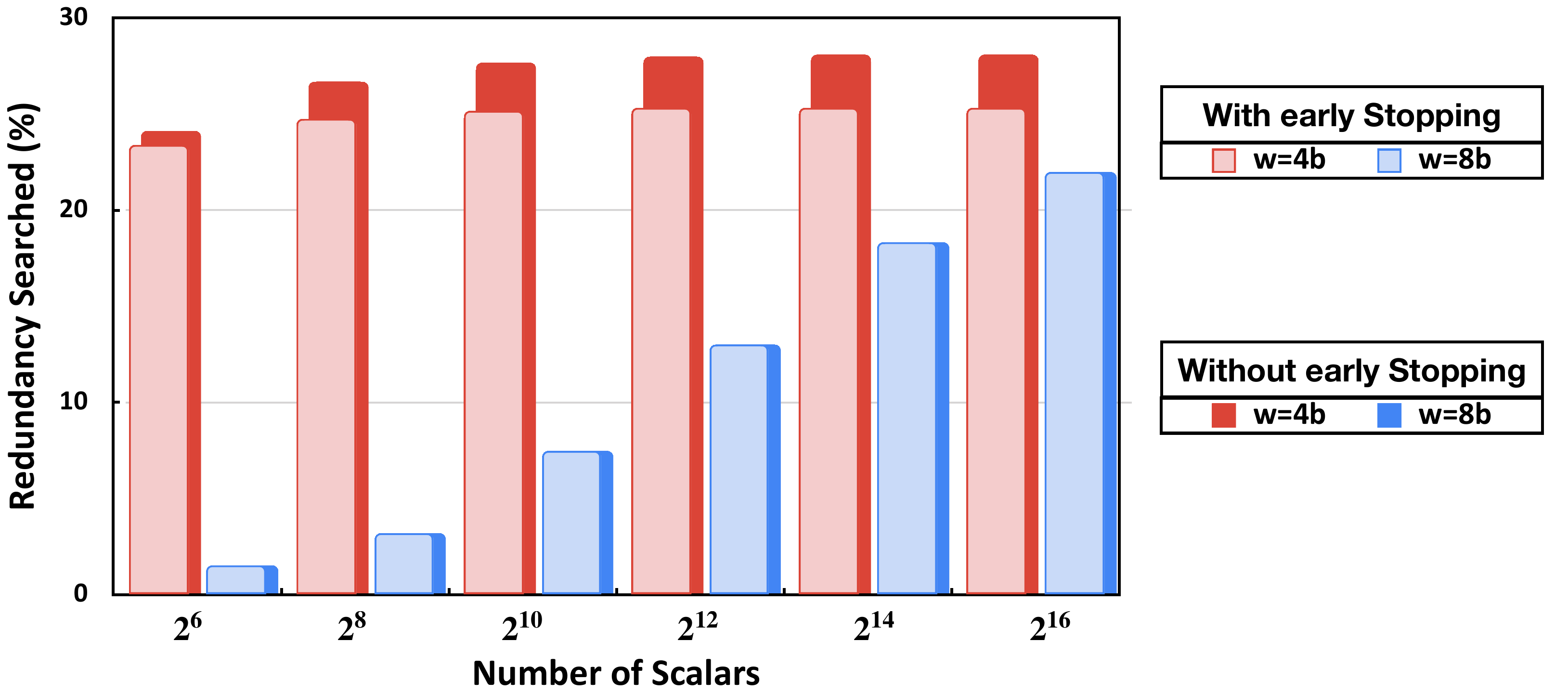}
    \caption{Redundancy vs. Number of Scalars for window size: 4b and 8b.} 
    \label{fig:w_N_pct}
  \end{subfigure}%
  \hfill
  \begin{subfigure}[t]{.5\linewidth}
    \centering
    \includegraphics[width=\linewidth]{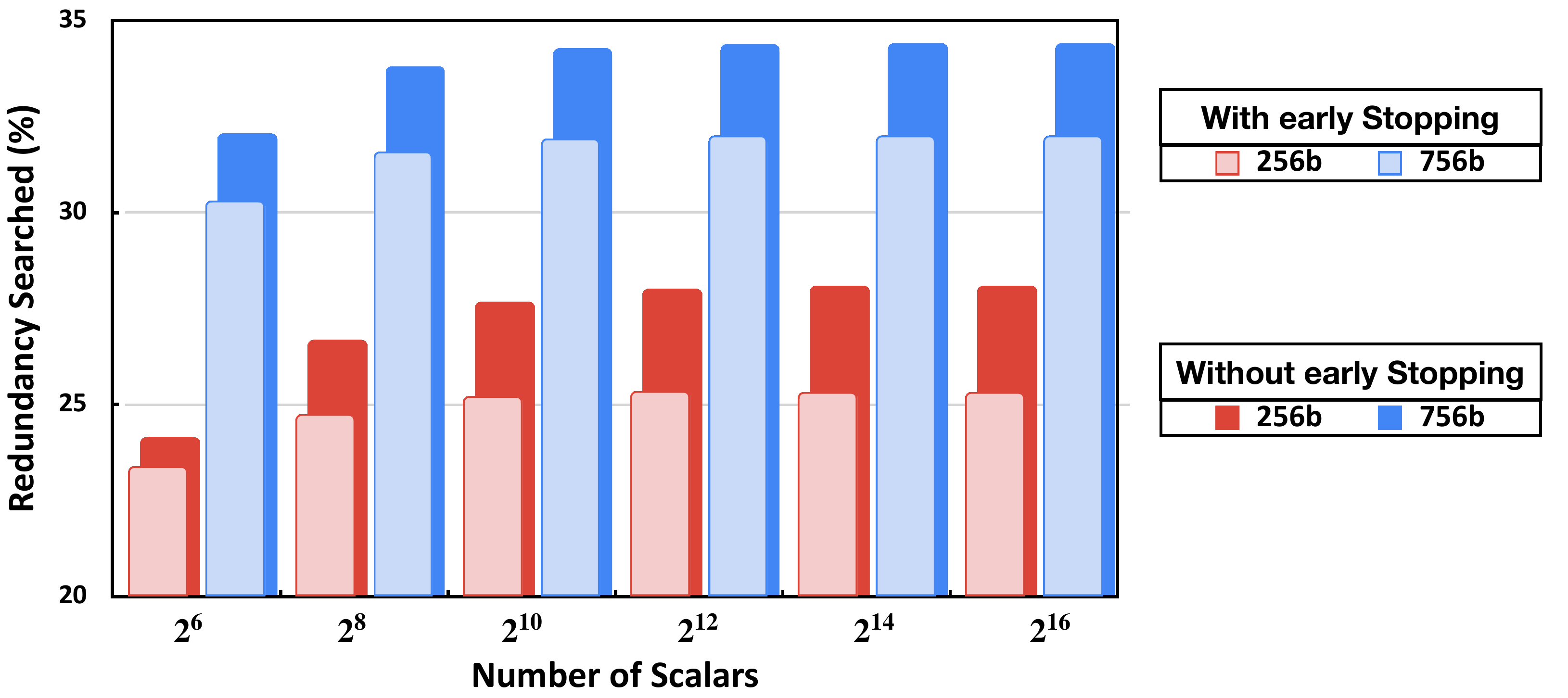}
    \caption{Redundancy vs. Number of Scalars for bitwidth: 256b and 756b.} 
    \label{fig:bitwidth_N_pct}
  \end{subfigure}
  \caption{SRE Algorithm: Redundancy vs. Number of Scalars for different window sizes and bitwidth.}
  \Description{}
  \label{fig:eval2}
\end{figure*}

\noindent \textbf{Window size \& redundancy searched.}
As mentioned in \Cref{sec:intro}, the number of operations required to compute MSM using the Pippenger’s bucket method heavily depends on the ratio between the number of scalars and the window size. While minimizing the number of operations is theoretically desirable, the optimal parameter choice ultimately depends on how the computation is implemented in software or hardware for performance acceleration. To account for this, we compare two window sizes in terms of the redundancy found during the point merging step for each setting. When the window size is 1 hexadecimal digit (i.e. 4 bits), the redundancy percentage saturates and the ratio remains within the same range. This behavior, previously discussed in the context of different approaches, is consistent across the evaluated methods.

However, when the window size increases to 2 hexadecimal digits (i.e. 8 bits), the redundancy percentage continues to grow with the number of scalars. In this case, applying the early stopping technique makes little difference. This is because, with an 8-bit window, many windows remain unmatched due to an insufficient number of scalars available for comparison. As a result, most comparisons fall below the threshold, allowing room for additional matches, and thus a potential increase in redundancy. As shown in \Cref{fig:w_N_pct}, the percentage of redundancy exhibits a steady upward trend without saturation. This suggests that with an even larger number of scalars, the benefits of using a larger window size become more pronounced. In our experiments, the maximum number of scalars tested was $2^{16}$, restricted by the exponentially increasing search time shown in \Cref{fig:sre_N_t}.

\noindent \textbf{Scalar bitwidth \& redundancy searched.}
As mentioned in \Cref{sec:intro} and the previous section, window size and bitwidth are key parameters that significantly influence the efficiency of Pippenger’s bucket method. In theory, the bitwidth is directly proportional to computational complexity; thus, regardless of the method used, increasing the bitwidth generally leads to higher computation costs and makes further reductions more difficult. However, our proposed algorithm exhibits the opposite behavior. When both the number of scalars and the window size are fixed, a larger bitwidth results in more windows per scalar, increasing the likelihood of finding matches. This makes the behavior of our algorithm diverge from that of Pippenger’s bucket method. In Pippenger’s approach, bitwidth does not affect how the algorithm operates, an increase in bitwidth simply translates into a linear increase in computation. This is evident in \Cref{fig:pippenger}, where the color coding represents different timestamps, and larger bitwidths correspond to more timestamps, indicating greater computational effort.

In contrast, our proposed SRE algorithm benefits from increased bitwidth. As shown in \Cref{fig:proposed}, where the color coding represents matches across all windows and scalars, a larger bitwidth expands the search space and creates more opportunities for redundancy detection. This highlights the advantage of applying our method in scenarios where Pippenger’s bucket method suffers from increased bitwidth overhead.

The experimental results are shown in \Cref{fig:bitwidth_N_pct}, where we evaluate the BLS12-377 and MNT6-753 curves, which have bitwidths of 256 bits and 756 bits, respectively, as listed in \Cref{tab:curve}. While the benefit of applying the early stopping technique remains consistent, we observe a fundamental increase in the percentage of redundancy found as the bitwidth increases. This higher redundancy percentage is maintained across different numbers of scalars, indicating the general applicability of our method. These findings are consistent with the observations and discussion presented earlier. Consequently, in scenarios where MSM requires higher security, and thus larger bitwidths, our proposed algorithm becomes even more advantageous.

\begin{figure*}[t] 
  \begin{subfigure}[t]{.502\linewidth}
    \centering
    \includegraphics[width=\linewidth]{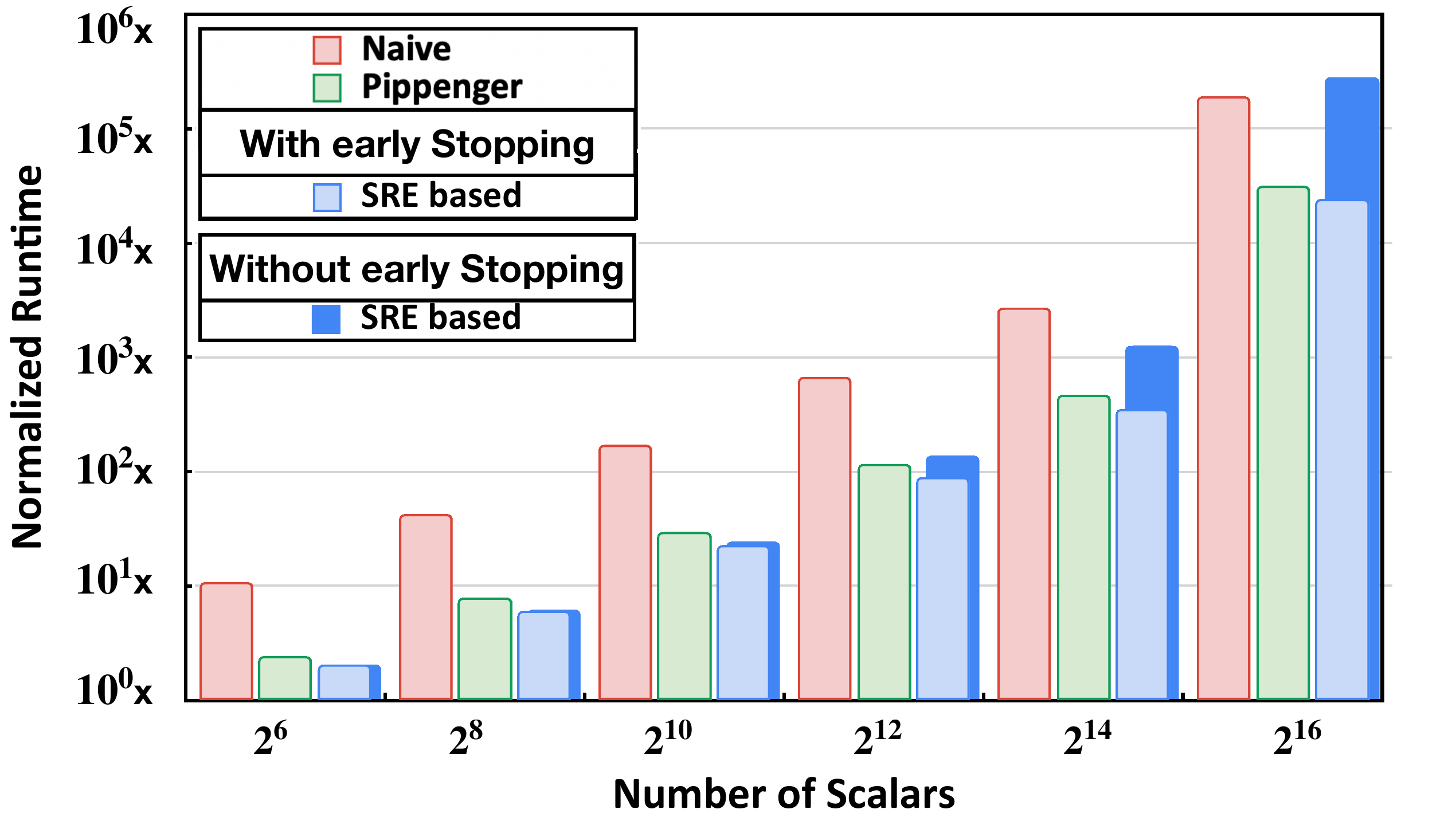}
    \caption{Runtime vs. Number of Scalars.} 
    \label{fig:msm_N_t}
  \end{subfigure}%
  \hfill
  \begin{subfigure}[t]{.49\linewidth}
    \centering
    \includegraphics[width=\linewidth]{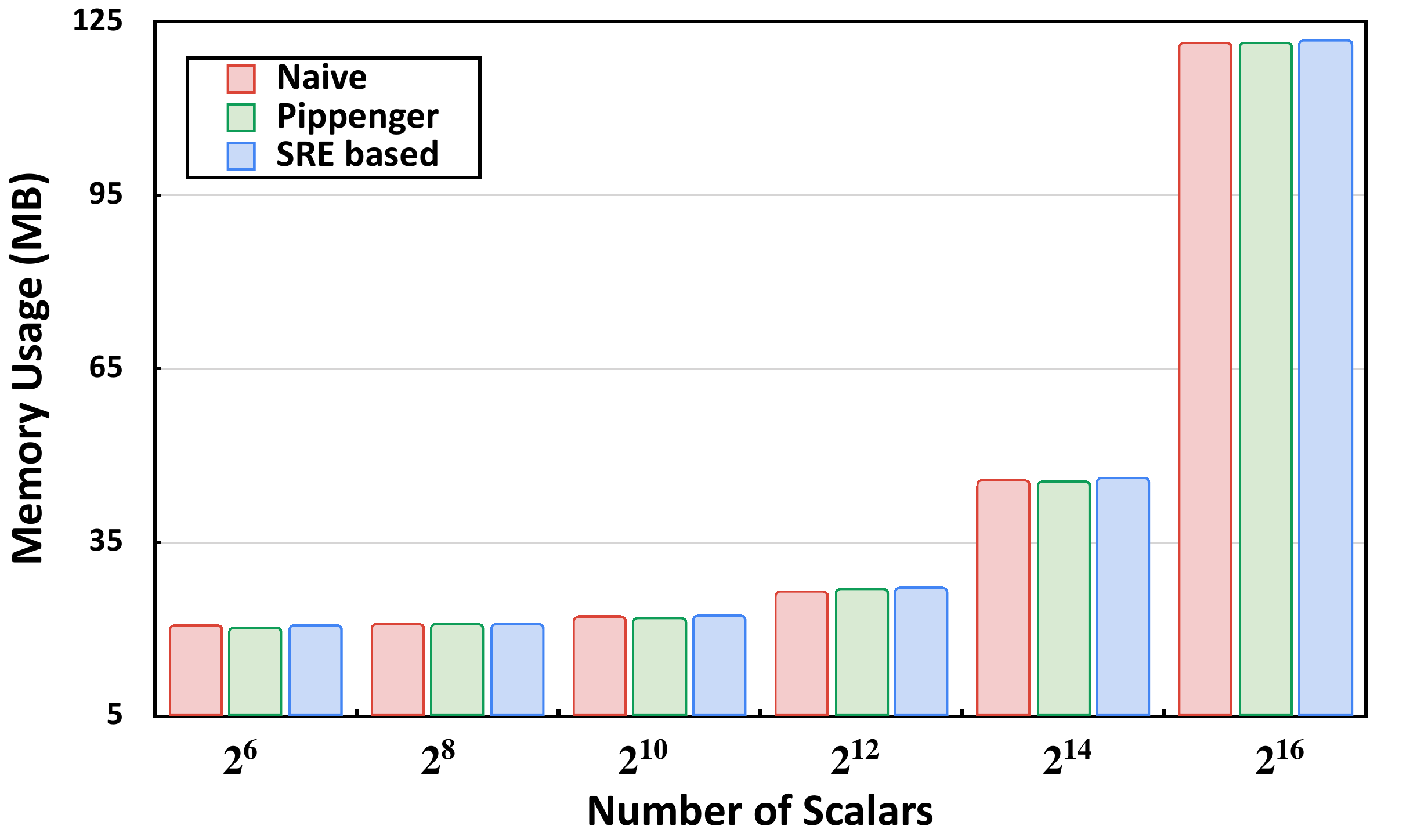}
    \caption{Memory Usage vs. Number of Scalars.} 
    \label{fig:msm_N_Mb}
  \end{subfigure}
  \caption{Multi-Scalar Multiplication: Runtime, Memory Usage vs. Number of Scalars for different algorithms.}
  \Description{}
  \label{fig:eval3}
\end{figure*}

\subsection{MSM Runtime \& Memory Usage Analysis}
In this section, we compare our proposed algorithm for computing MSM with the naive method and the Pippenger’s bucket method. We integrate our SRE algorithm into MSM computation and analyze the actual benefits and tradeoffs between the additional search cost and the redundancy reduction it enables. These tradeoffs are reflected in the runtime and memory usage of each process. Our experiments were conducted using varying numbers of scalars under a vanilla array based approach of our SRE algorithm, with a window size of 4 bits and the BLS12-377 curve. We computed MSM using all three methods and compared the result for functional correctness before conducting experiments on runtime and memory usage. 

\begin{table}[b]
  \centering
  \caption{Proposed algorithm speedup comparison \& SRE search latency to proposed algorithm ratio. } \label{tab:speedup}
  \begin{threeparttable}
    \begin{tabular}{c|ccc|cc}
      \toprule
      N        & Pippenger & Proposed & Speedup & SRE   & Search ratio\\
      \midrule
      $2^{6}$  & 2.43      & 2.04     & 19.04\% & 0.09  & 4.44\%      \\
      $2^{8}$  & 7.91      & 6.14     & 28.64\% & 0.21  & 3.47\%      \\
      $2^{10}$ & 29.75     & 22.66    & 31.33\% & 0.69  & 3.04\%      \\
      $2^{12}$ & 117.34    & 88.98    & 31.87\% & 2.52  & 2.84\%      \\
      $2^{14}$ & 489.34    & 381.19   & 28.37\% & 10.07 & 2.64\%      \\
      $2^{16}$ & 33385.32  & 25307.95 & 31.92\% & 40.07 & 2.73\%      \\
      \bottomrule
    \end{tabular}
    \begin{tablenotes}
      \scriptsize
      \item[*] Normalized runtime to proposed method in $2^4$ number of scalars.  
      \item[*] Consider Pippenger's algorithm as baseline for speedup.  
    \end{tablenotes} 
  \end{threeparttable}
\end{table}

\noindent \textbf{Runtime.}
As shown in \Cref{fig:msm_N_t}, we conducted experiments on MSM computation using three methods: the naive method, Pippenger’s bucket method, and our proposed SRE-based method. The runtime is normalized to that of our proposed algorithm with $2^4$ scalars. The normalized runtime is presented on a logarithmic scale, where results with and without the early stopping technique are overlaid to highlight the additional search cost incurred in exchange for detecting more redundancy. Since our method builds upon and improves Pippenger’s bucket method, the actual runtimes are of the same order of magnitude. In contrast, the naive method exhibits runtimes approximately an order of magnitude higher, which highlights the overall efficiency advantage of Pippenger's bucket method and our proposed SRE-based method.

To better evaluate the performance of our proposed algorithm, we calculate the percentage speedup, as shown in \Cref{tab:speedup}, using Pippenger’s bucket method as the baseline. Compared with the values in \Cref{fig:sre_N_pct}, the speedup shows a strong correlation with the percentage of redundancy identified. The observed speedup can be attributed to two key factors: the reduction in runtime due to eliminated redundancy, and the overhead introduced by the additional search operations. We define the search ratio as the normalized runtime of the SRE algorithm (using the vanilla array-based approach) with respect to the total MSM runtime in our proposed method. This ratio ranges between 2–4\%, which is a reasonable tradeoff given the amount of redundancy detected. We also compared the runtime of executing the theoretical number of operations with the actual MSM runtime. Since PADDs and PDBLs dominate the computation, their count directly impacts performance. Considering all contributing factors, the search time introduced by our SRE algorithm is relatively small but leads to a significant reduction in MSM runtime. Compared to Pippenger’s bucket method, our approach achieves approximately a 30\% speedup.

\noindent \textbf{Memory usage.}
The memory usage of each algorithm across different numbers of scalars is shown in \Cref{fig:msm_N_Mb}. The reported memory usage accounts for all stored scalars and points, as well as any intermediate values used during computation. When the number of scalars is small, the difference in memory usage between algorithms remains under 5\%, indicating that the base program size dominates overall memory consumption rather than the data size itself. 

As the number of scalars increases, the impact on memory usage becomes more significant, growing at an exponential rate. However, the relative memory usage between different algorithms remains nearly identical. In conjunction with \Cref{tab:storage}, which identifies scalars and points as the main sources of memory demand, we observe that our proposed algorithm introduces minimal overhead compared to the entire computational size. This overhead is substantially smaller than that of precomputation-based methods, where the majority of memory is consumed by large lookup tables, often several times the size of the MSM computation itself.

\section{Future Work} \label{sec:future}
Our key observation, merging multiple windows of two scalars in the Pippenger Algorithm, opens a new research direction focused on maximizing possible merges. While our current approach significantly reduces operations and accelerates MSM, the search algorithms remain simple. We aim to develop more efficient algorithms that maximize merges with minimal comparisons. The observation also implies that EC point operations can be further reduced when multiple scalars share common windows, not just pairs. Therefore, extending search algorithms to handle more than two scalars is a promising avenue for future work.

The problem of finding maximum merges in scalars is related to similarity search problems in statistics and machine learning, such as k-nearest neighbors (KNN), hierarchical navigable small world (HNSW), and locality-sensitive hashing (LSH). However, our case differs in that we aim to exhaustively find all possible merges, repeating the process until no further merges remain or the search time outweighs the computation cost. This requires frequent querying and dynamic updates to the dataset, which existing algorithms handle poorly. While they offer fast queries, their update times are inefficient. Therefore, a new algorithm tailored to our needs is essential to improve search efficiency and potentially reduce overall computation time. 
\section{Conclusion} \label{sec:conclusion}
MSM in ECC is a fundamental cryptographic operations that has various applications, from aggregated signatures to polynomial commitments in ZKPs. Due to the complexity of this computation, it is the dominant factor in runtime. In this work, we presented a unique SRE algorithm for eliminating the redundant operations in the traditional Pippenger's bucket method and proposed our MSM algorithm integration with our search algorithm. We provide three approaches for SRE algorithm with their tradeoffs in runtime and resources used. Experimental result shows a small extra search time cost of 3\% for redundancy with more than 25\% of redundancy operations eliminated and saved thanks to our algorithm. We presented a novel method in tackling MSM and practical applications with our experimental results shown.


\bibliographystyle{abbrvnat}
\bibliography{misc/refs}



\end{document}